\documentclass[journal]{IEEEtran}

\usepackage[utf8]{inputenc}
\usepackage{color}
\usepackage{array}
\usepackage{verbatim}
\usepackage{float}
\usepackage{amsmath}
\usepackage{amsthm}
\usepackage{amssymb}
\usepackage{graphicx}
\usepackage{longtable}
\usepackage{multirow}
\usepackage[unicode=true,
bookmarks=false,
breaklinks=false,pdfborder={0 0 1},colorlinks=false]
{hyperref}
\hypersetup{
	colorlinks,bookmarksopen,bookmarksnumbered,citecolor=blue,urlcolor=blue}
\usepackage{cite}

\makeatletter
\let\oldforeign@language\foreign@language
\DeclareRobustCommand{\foreign@language}[1]{%
	\lowercase{\oldforeign@language{#1}}}

\let\oldforeign@language\foreign@language
\DeclareRobustCommand{\foreign@language}[1]{%
	\lowercase{\oldforeign@language{#1}}}

\ifCLASSINFOpdf
\else
\fi

\newcommand{\MYfooter}{\smash{
		\hfil\parbox[t][\height][t]{\textwidth}{\centering
			\thepage}\hfil\hbox{}}}

\makeatletter

\def\ps@IEEEtitlepagestyle{%
	\def\@oddhead{\parbox[t][\height][t]{\textwidth}{\centering \scriptsize
			Personal use of this material is permitted. Permission from the author(s) and/or copyright holder(s), must be obtained for all other uses. Please contact us and provide details if you believe this document breaches copyrights.\\
			\noindent\makebox[\linewidth]{}
		}\hfil\hbox{}}%
	\def\@evenhead{\scriptsize\thepage \hfil \leftmark\mbox{}}%
	\def\@oddfoot{\parbox[t][\height][l]{\textwidth}{
			\vspace{-20pt}{\rule{\textwidth}{0.4pt}}\\ \footnotesize\underline{To cite this article:}
			{\bf{\textcolor{red}{H. A. Hashim, and F. L. Lewis, "Nonlinear Stochastic Estimators on the Special Euclidean Group SE(3) Using Uncertain IMU and Vision Measurements," IEEE Transactions on Systems, Man, and Cybernetics: Systems, vol. 51, no. 12, pp. 7587-7600, 2021.}}} doi: \href{https://doi.org/10.1109/TSMC.2020.2980184}{10.1109/TSMC.2020.2980184}\\
			\noindent\makebox[\linewidth]
		}\hfil\hbox{}}%
	\def\@evenfoot{\MYfooter}}

\makeatother
\newtheorem{defn}{Definition}

\newtheorem{lem}{Lemma}

\newtheorem{thm}{Theorem}
\newtheorem{rem}{Remark}

\newtheorem{assum}{Assumption}

\begin{document}
	\bstctlcite{IEEEexample:BSTcontrol}

\title{Nonlinear Stochastic Estimators on the Special Euclidean Group SE(3) using Uncertain IMU and Vision Measurements}

\author{Hashim A. Hashim$^*$\IEEEmembership{~Member, IEEE} and Frank L. Lewis\IEEEmembership{~Fellow, IEEE}
	\thanks{This work was supported in part by Thompson Rivers University Internal research fund \# 102315.}
	\thanks{$^*$Corresponding author, H. A. Hashim is with the Department of Engineering and Applied Science, Thompson Rivers University, Kamloops, British Columbia, Canada, V2C-0C8, e-mail: hhashim@tru.ca.}
	\thanks{F. L. Lewis is with the Department of Electrical and Computer Engineering, UTA Research Institute, The University of Texas at Arlington 7300 Jack Newell Blvd. S, Ft. Worth, Texas 76118, e-mail: lewis@uta.edu.}
	
}


\markboth{}{Hashim \MakeLowercase{\textit{et al.}}: Nonlinear Pose Filters on SE(3) with Guaranteed Transient and Steady-state Performance}

\maketitle

\begin{abstract}
Two novel robust nonlinear stochastic full pose (\textit{i.e}, attitude
and position) estimators on the Special Euclidean Group $\mathbb{SE}\left(3\right)$
are proposed using the available uncertain measurements. The resulting
estimators utilize the basic structure of the deterministic pose estimators
adopting it to the stochastic sense. The proposed estimators for six
degrees of freedom (DOF) pose estimations consider the group velocity
vectors to be contaminated with constant bias and  Gaussian random
noise, unlike nonlinear deterministic pose estimators which disregard
the noise component in the estimator derivations. The proposed estimators
ensure that the closed loop error signals are semi-globally uniformly
ultimately bounded in mean square. The equivalent quaternion representation and complete implementation
steps of the proposed filters are presented. The efficiency and robustness
of the proposed estimators are demonstrated by the numerical results
which test the estimators against high levels of noise and bias associated
with the group velocity and body-frame measurements and large initialization
error. 
\end{abstract}


\IEEEpeerreviewmaketitle{}

\section{Introduction}

\IEEEPARstart{L}{andmark-based} navigation is an integral part of robotics and control
applications due to its ability to identify the pose (\textit{i.e}.,
attitude and position) of a rigid-body in three-dimensional (3D) space.
Applications requiring accurate 3D pose information include, but are
not limited to, sensor calibration \cite{rehbinder2003pose}, manipulation
and registration \cite{srivatsan2016estimating}, and tracking control
of autonomous vehicles \cite{kwon2007particle,vasconcelos2010nonlinear,mohamed2019filters}.
The orientation of a rigid-body, also known as attitude, cannot be
measured directly, instead, it has to be reconstructed using one of
the following methods \cite{hashim2020AtiitudeSurvey}: static reconstruction
\cite{shuster1981three,markley1988attitude}, Gaussian filter estimation
\cite{lefferts1982kalman,markley2003attitude,choukroun2006novel},
or nonlinear-based estimators \cite{mahony2008nonlinear,liu2018complementary,hashim2018SO3Stochastic,hashim2018Conf1}.
The static methods of attitude reconstructions such as QUEST \cite{shuster1981three}
or singular value decomposition (SVD) \cite{markley1988attitude}
utilize two or more known non-collinear observations in the inertial-frame
and their sensor measurements in the body-frame. Nonetheless, it is
worth noting that sensor measurements are vulnerable to bias and noise
components causing the algorithms in \cite{shuster1981three,markley1988attitude}
to produce poor results, especially if the vehicle is equipped with
low-cost inertial measurement units (IMU).

Conventionally, the attitude estimation problem is predominantly addressed
using Gaussian filters, for instance, Kalman filter (KF) \cite{choukroun2006novel},
extended KF (EKF) \cite{lefferts1982kalman}, multiplicative EKF (MEKF)
\cite{markley2003attitude}, and for good survey of Gaussian attitude
estimator visit \cite{hashim2018SO3Stochastic,mohamed2019filters}.
Gaussian filters generate reliable attitude estimates when the rigid-body
is equipped with high quality measurement units. Despite all the benefits
offered by Gaussian filters, high quality measurement units have multiple
disadvantages, namely large size, heavy weight, and high cost. The
recent rise of micro-elector-mechanical systems (MEMS) allowed for
development of IMU, which are relatively inexpensive, small in size,
and light-weight. However, the output of the low-cost IMU is contaminated
with noise resulting in unsatisfactory performance of Gaussian attitude
filters \cite{hashim2018SO3Stochastic,mohamed2019filters,hashim2019SO3Wiley}. Consequently,
numerous nonlinear complementary estimators evolved directly on the
Special Orthogonal Group $\mathbb{SO}\left(3\right)$ have been proposed,
for example \cite{mahony2008nonlinear,liu2018complementary,zlotnik2017nonlinear,hashim2018SO3Stochastic,hashim2018Conf1,hashim2019SO3Wiley}.
Nonlinear complementary estimators have been proven to outperform
Gaussian filters in multiple respects, namely, 1) nonlinear complementary
estimator design accounts for the nonlinear nature of the attitude
problem, 2) their derivation and representation is considerably simpler,
3) they require less computational cost, and 4) show better tracking
performance \cite{mahony2008nonlinear,hashim2018SO3Stochastic,mohamed2019filters}.
Pose estimation is also best approached in nonlinear sense (on the
Lie group of the Special Euclidean Group $\mathbb{SE}\left(3\right)$),
since nonlinear attitude estimation is an integral component of pose
estimation. 

The structure of nonlinear pose estimators developed on $\mathbb{SE}\left(3\right)$
relies on angular and translational velocity measurements, vector
measurements, landmark(s) measurements, and estimates of the uncertain
components associated with the velocity measurements (for example
\cite{hashim2019SE3Det,mohamed2019filters,rehbinder2003pose,vasconcelos2010nonlinear,baldwin2007complementary,baldwin2009nonlinear,hua2011observer}).
{} With the aim of improving the convergence behavior, several nonlinear
deterministic pose estimators have been proposed \cite{rehbinder2003pose,vasconcelos2010nonlinear,baldwin2007complementary,baldwin2009nonlinear,hua2011observer,dominguez2017simultaneous,hua2017riccati}.
An early implementation of nonlinear deterministic pose estimator
with an inertial vision system was introduced in \cite{rehbinder2003pose}.
It was followed by a semi-direct deterministic pose estimator on $\mathbb{SE}\left(3\right)$
which required pose reconstruction \cite{baldwin2007complementary}.
The work in \cite{baldwin2007complementary} has been modified to
obtain a direct deterministic pose estimator on $\mathbb{SE}\left(3\right)$
\cite{hua2011observer} which utilizes the measurements directly,
thus obviating the necessity for pose reconstruction. The noteworthy
feature of the nonlinear deterministic pose estimators in \cite{rehbinder2003pose,vasconcelos2010nonlinear,baldwin2007complementary,baldwin2009nonlinear,hua2011observer,dominguez2017simultaneous,hua2017riccati,hashim2019SE3Det}
is the guarantee of the almost global asymptotic stability of the
pose error achieved by disregarding the random noise attached to the
group velocity vector. However, it is common for the group velocity
vector measurements to be contaminated with constant bias and random
noise. Bias and noise have the potential to compromise the estimation
process and lead to poor results, in particular, if the vehicle is
fitted with low-cost inertial vision system which includes an IMU
module and a vision system. Several nonlinear stochastic estimators
have been developed that addressed the sensitivity to measurement
noise, for instance, \cite{cao2019adaptive}, and \cite{luo2019latent}
and bias estimation problem \cite{geng2018target}.%
{} 

Concluding the introductory overview of the pose problem, it is important
to emphasize two critical considerations. Firstly, the pose problem
is naturally nonlinear on the Lie group of $\mathbb{SE}\left(3\right)$.
Secondly, the group velocity vectors are not only corrupted with constant
bias but also with random noise. The two nonlinear stochastic pose
estimators on the Lie group of $\mathbb{SE}\left(3\right)$ proposed
in this paper take into account the above-mentioned considerations
and use data extracted by an IMU module and a vision system. In case
when the group velocity vector is contaminated with constant bias
and Gaussian random noise, the advantages of the proposed estimators
are as follows: 1) The closed loop error signals are guaranteed to
be almost semi-globally uniformly ultimately bounded in mean square.
2) The noise contamination of the estimator dynamics is minimized.
3) Unlike previously proposed nonlinear deterministic estimators,
the proposed stochastic estimators produce reliable pose estimate
and successfully handle irregular behavior of the measurement noise
as well as large initialization error. 

The rest of the paper is organized as follows: Section \ref{sec:Preliminaries-and-Math}
introduces $\mathbb{SO}\left(3\right)$ and $\mathbb{SE}\left(3\right)$
preliminaries and mathematical notation. In Section \ref{sec:SE3_Problem-Formulation}
the pose problem is presented in stochastic sense. Section \ref{sec:SE3_Stochastic-Complementary-Filters}
proposes two nonlinear stochastic pose estimators on $\mathbb{SE}\left(3\right)$
including related stability analysis. Section \ref{sec:SE3_Simulations}
illustrates the effectiveness and robustness of the proposed estimation
schemes. Finally, Section \ref{sec:SE3_Conclusion} concludes the
work. 

\section{Preliminaries and Math Notation \label{sec:Preliminaries-and-Math}}

Throughout the paper, the set of non-negative real numbers, real $n$-dimensional
space, and real $n\times m$ dimensional space are referred to as
$\mathbb{R}_{+}$, $\mathbb{R}^{n}$, and $\mathbb{R}^{n\times m}$,
respectively. For any $x\in\mathbb{R}^{n}$, $\left[x\right]_{{\rm D}}$
denotes a diagonal matrix of $x$ and $^{\top}$ denotes a transpose
of a component. $\left\Vert x\right\Vert =\sqrt{x^{\top}x}$ stands
for the Euclidean norm of $x\in\mathbb{R}^{n}$. The $n$-by-$n$
identity matrix is referred to as $\mathbf{I}_{n}$. $\mathcal{C}^{n}$
stands for the $n$th continuous partial derivative of a continuous
function. $\mathcal{K}$ describes a set of continuous and strictly
increasing functions which follows $\gamma:\mathbb{R}_{+}\rightarrow\mathbb{R}_{+}$
and is zero only at the origin. $\mathcal{K}_{\infty}$, despite being
a class $\mathcal{K}$ function, is unbounded. ${\rm Tr}\left\{ \cdot\right\} $,
$\mathbb{P}\left\{ \cdot\right\} $, and $\mathbb{E}\left[\cdot\right]$
denote trace, probability, and an expected value of a component, respectively.
$\left\{ \mathcal{B}\right\} $ denotes the body-frame and $\left\{ \mathcal{I}\right\} $
denotes the inertial-frame.

The orthogonal group $\mathbb{O}\left(3\right)$ is a Lie group and
a subgroup of the 3-dimensional general linear group, characterized
by smooth multiplication and inversion and defined by
\[
\mathbb{O}\left(3\right)=\{\left.\boldsymbol{M}\in\mathbb{R}^{3\times3}\right|\boldsymbol{M}^{\top}\boldsymbol{M}=\boldsymbol{M}\boldsymbol{M}^{\top}=\mathbf{I}_{3}\}
\]
where $\mathbf{I}_{3}\in\mathbb{R}^{3\times3}$ is the identity matrix.
The Special Orthogonal Group $\mathbb{SO}\left(3\right)$ is a subgroup
of $\mathbb{O}\left(3\right)$ and is given by
\[
\mathbb{SO}\left(3\right)=\{\left.R\in\mathbb{R}^{3\times3}\right|RR^{\top}=R^{\top}R=\mathbf{I}_{3}\text{, }{\rm det}\left(R\right)=+1\}
\]
where ${\rm det\left(\cdot\right)}$ is a determinant of a matrix,
and $R\in\mathbb{SO}\left(3\right)$ describes the orientation, commonly
known as attitude, of a rigid-body in the body-frame relative to the
inertial-frame in 3D space. The Special Euclidean Group $\mathbb{SE}\left(3\right)$
is a subset of the affine group defined by
\[
\mathbb{SE}\left(3\right)=\left\{ \left.\boldsymbol{T}=\left[\begin{array}{cc}
R & P\\
\mathbf{0}_{3\times1}^{\top} & 1
\end{array}\right]\in\mathbb{R}^{4\times4}\right|R\in\mathbb{SO}\left(3\right),P\in\mathbb{R}^{3}\right\} 
\]
where $\boldsymbol{T}\in\mathbb{SE}\left(3\right)$ is a homogeneous
transformation matrix that describes the pose of a rigid-body in 3D
space, while $P\in\mathbb{R}^{3}$ stands for position, $R\in\mathbb{SO}\left(3\right)$.
The Lie-algebra of the group $\mathbb{SO}\left(3\right)$ is termed
$\mathfrak{so}\left(3\right)$ and expressed as
\[
\mathfrak{so}\left(3\right)=\{\left.\left[x\right]_{\times}\in\mathbb{R}^{3\times3}\right|\left[x\right]_{\times}^{\top}=-\left[x\right]_{\times}\}
\]
with $\left[x\right]_{\times}$ being a skew symmetric matrix such
that the map $\left[\cdot\right]_{\times}:\mathbb{R}^{3}\rightarrow\mathfrak{so}\left(3\right)$
is given by
\[
\left[x\right]_{\times}=\left[\begin{array}{ccc}
0 & -x_{3} & x_{2}\\
x_{3} & 0 & -x_{1}\\
-x_{2} & x_{1} & 0
\end{array}\right]\in\mathfrak{so}\left(3\right),\hspace{1em}x=\left[\begin{array}{c}
x_{1}\\
x_{2}\\
x_{3}
\end{array}\right]
\]
Define $\left[x\right]_{\times}y=x\times y$ where $\times$ denotes
the cross product for all $x,y\in\mathbb{R}^{3}$. $\mathfrak{se}\left(3\right)$
is a Lie-algebra of $\mathbb{SE}\left(3\right)$ such that{\small{}
	\[
	\mathfrak{se}\left(3\right)=\left\{ \left.\left[\mathcal{Y}\right]_{\wedge}\in\mathbb{R}^{4\times4}\right|\exists y_{1},y_{2}\in\mathbb{R}^{3}:\left[\mathcal{Y}\right]_{\wedge}=\left[\begin{array}{cc}
	\left[y_{1}\right]_{\times} & y_{2}\\
	\mathbf{0}_{3\times1}^{\top} & 0
	\end{array}\right]\right\} 
	\]
}where the wedge map $\left[\cdot\right]_{\wedge}:\mathbb{R}^{6}\rightarrow\mathfrak{se}\left(3\right)$
is defined by
\[
\left[\mathcal{Y}\right]_{\wedge}=\left[\begin{array}{cc}
\left[y_{1}\right]_{\times} & y_{2}\\
\mathbf{0}_{3\times1}^{\top} & 0
\end{array}\right]\in\mathfrak{se}\left(3\right),\hspace{1em}\mathcal{Y}=\left[\begin{array}{c}
y_{1}\\
y_{2}
\end{array}\right]\in\mathbb{R}^{6}
\]
On the other side, the inverse of $\left[\cdot\right]_{\times}$ is
$\mathbf{vex}:\mathfrak{so}\left(3\right)\rightarrow\mathbb{R}^{3}$,
such that for $\alpha\in\mathbb{R}^{3}$ one has
\begin{equation}
\mathbf{vex}(\left[\alpha\right]_{\times})=\alpha\label{eq:SE3STCH_VEX}
\end{equation}
The anti-symmetric projection on the Lie-algebra of $\mathfrak{so}\left(3\right)$
is defined by $\boldsymbol{\mathcal{P}}_{a}$ and its mapping follows
$\boldsymbol{\mathcal{P}}_{a}:\mathbb{R}^{3\times3}\rightarrow\mathfrak{so}\left(3\right)$
such that
\begin{equation}
\boldsymbol{\mathcal{P}}_{a}\left(\boldsymbol{M}\right)=\frac{1}{2}\left(\boldsymbol{M}-\boldsymbol{M}^{\top}\right)\in\mathfrak{so}\left(3\right),\,\boldsymbol{M}\in\mathbb{R}^{3\times3}\label{eq:SE3STCH_Pa}
\end{equation}
Let $\boldsymbol{\Upsilon}_{a}\left(\cdot\right)$ represent the composition
mapping $\boldsymbol{\Upsilon}_{a}=\mathbf{vex}\circ\boldsymbol{\mathcal{P}}_{a}$.
Accordingly, for $\boldsymbol{M}\in\mathbb{R}^{3\times3}$ one has
\begin{equation}
\boldsymbol{\Upsilon}_{a}\left(\boldsymbol{M}\right)=\mathbf{vex}(\boldsymbol{\mathcal{P}}_{a}\left(\boldsymbol{M}\right))\in\mathbb{R}^{3}\label{eq:SE3STCH_VEX_a}
\end{equation}
The normalized Euclidean distance of the attitude matrix $R\in\mathbb{SO}\left(3\right)$
is defined as follows
\begin{equation}
\left\Vert R\right\Vert _{{\rm I}}=\frac{1}{4}{\rm Tr}\left\{ \mathbf{I}_{3}-R\right\} \in\left[0,1\right]\label{eq:SE3STCH_Ecul_Dist}
\end{equation}
The orientation of any rigid-body can be established knowing its angle
of rotation $\alpha\in\mathbb{R}$ about the unit-axis $u\in\mathbb{R}^{3}$
in the sphere $\mathbb{S}^{2}$. This method of attitude representation
is generally termed to as angle-axis parameterization \cite{shuster1993survey}.
The mapping of angle-axis parameterization to $\mathbb{SO}\left(3\right)$
is defined by $\mathcal{R}_{\alpha}:\mathbb{R}\times\mathbb{R}^{3}\rightarrow\mathbb{SO}\left(3\right)$
with
\begin{equation}
\mathcal{R}_{\alpha}\left(\alpha,u\right)=\mathbf{I}_{3}+\sin\left(\alpha\right)\left[u\right]_{\times}+\left(1-\cos\left(\alpha\right)\right)\left[u\right]_{\times}^{2}\label{eq:SE3STCH_att_ang}
\end{equation}
For $x,y\in{\rm \mathbb{R}}^{3}$, $R\in\mathbb{SO}\left(3\right)$,
$A\in\mathbb{R}^{3\times3}$, and $B=B^{\top}\in\mathbb{R}^{3\times3}$
the following mathematical identities will be used in the subsequent
derivations 
\begin{align}
\left[x\times y\right]_{\times}= & yx^{\top}-xy^{\top}\label{eq:SE3STCH_Identity1}\\
\left[Rx\right]_{\times}= & R\left[x\right]_{\times}R^{\top}\label{eq:SE3STCH_Identity2}\\
\left[x\right]_{\times}^{2}= & -x^{\top}x\mathbf{I}_{3}+xx^{\top}\label{eq:SE3STCH_Identity3}\\
B\left[x\right]_{\times}+\left[x\right]_{\times}B= & {\rm Tr}\left\{ B\right\} \left[x\right]_{\times}-\left[Bx\right]_{\times}\label{eq:SE3STCH_Identity4}\\
{\rm Tr}\left\{ B\left[x\right]_{\times}\right\} = & 0\label{eq:SE3STCH_Identity5}
\end{align}
{\small{}
	\begin{align}
	{\rm Tr}\left\{ A\left[x\right]_{\times}\right\} = & {\rm Tr}\left\{ \boldsymbol{\mathcal{P}}_{a}\left(A\right)\left[x\right]_{\times}\right\} =-2\mathbf{vex}\left(\boldsymbol{\mathcal{P}}_{a}\left(A\right)\right)^{\top}x\label{eq:SE3STCH_Identity6}
	\end{align}
}{\small\par}

\section{Problem Formulation\label{sec:SE3_Problem-Formulation}}

The pose estimation problem involves a set of vector measurements
made with respect to the inertial- and body-frames of reference. In
this section the pose problem is defined and the associated measurements
are presented. 

\begin{figure}[h]
	\centering{}\includegraphics[scale=0.43]{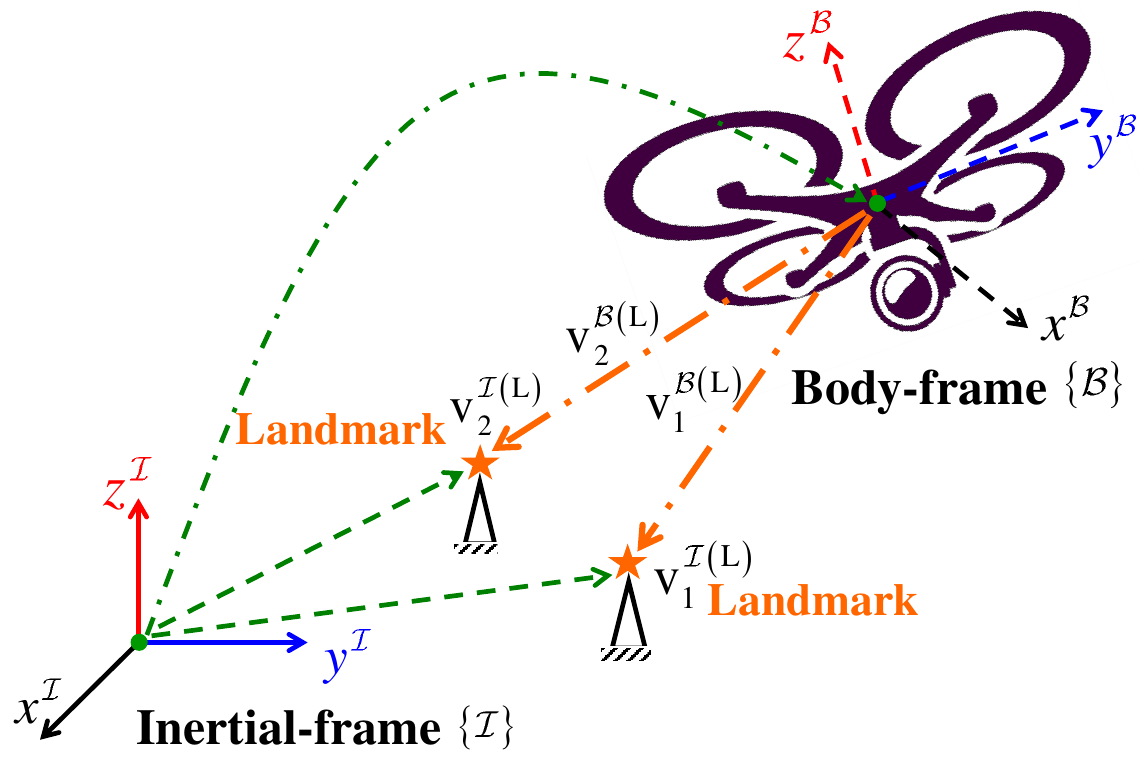}\caption{Pose estimation problem of a rigid-body in 3D space.}
	\label{fig:SE3PPF_1}
\end{figure}

Attitude and position are the two elements necessary to describe the
pose of a rigid-body in 3D space. Therefore, producing reliable estimates
of these two elements is at the core of this work. The orientation
of a rigid-body is termed attitude $R\in\mathbb{SO}\left(3\right)$
and defines the body orientation in the body-frame relative to the
inertial-frame $R\in\left\{ \mathcal{B}\right\} $. The translation
of the rigid-body is represented by $P\in\mathbb{R}^{3}$ where $P$
is defined relative to the inertial-frame $P\in\left\{ \mathcal{I}\right\} $.
Fig. \ref{fig:SE3PPF_1} illustrates the pose estimation problem of
a rigid-body in 3D space. Thus, the pose of a rigid-body is represented
by the following homogeneous transformation matrix $\boldsymbol{T}\in\mathbb{SE}\left(3\right)$:
\begin{equation}
\boldsymbol{T}=\left[\begin{array}{cc}
R & P\\
\mathbf{0}_{3\times1}^{\top} & 1
\end{array}\right]\label{eq:SE3PPF_T_matrix2}
\end{equation}

For clarity, the superscripts $\mathcal{B}$ and $\mathcal{I}$ are
used to differentiate components of body-frame and inertial-frame,
respectively. From one side, the attitude can be extracted given the
availability of $N_{{\rm R}}$ known non-collinear observations in
the inertial-frame and their measurements in the body-frame. The body-frame
measurements can be obtained, for instance, by low cost IMU, and the
$i$th measurement can be represented by
\[
\left[\begin{array}{c}
{\rm v}_{i}^{\mathcal{B}\left({\rm R}\right)}\\
0
\end{array}\right]=\boldsymbol{T}^{-1}\left[\begin{array}{c}
{\rm v}_{i}^{\mathcal{I}\left({\rm R}\right)}\\
0
\end{array}\right]+\left[\begin{array}{c}
{\rm b}_{i}^{\mathcal{B}\left({\rm R}\right)}\\
0
\end{array}\right]+\left[\begin{array}{c}
\omega_{i}^{\mathcal{B}\left({\rm R}\right)}\\
0
\end{array}\right]
\]
More simply put, 
\begin{equation}
{\rm v}_{i}^{\mathcal{B}\left({\rm R}\right)}=R^{\top}{\rm v}_{i}^{\mathcal{I}\left({\rm R}\right)}+{\rm b}_{i}^{\mathcal{B}\left({\rm R}\right)}+\omega_{i}^{\mathcal{B}\left({\rm R}\right)}\label{eq:SE3STCH_Vect_R}
\end{equation}
where ${\rm v}_{i}^{\mathcal{I}\left({\rm R}\right)}$, ${\rm b}_{i}^{\mathcal{B}\left({\rm R}\right)}$,
and $\omega_{i}^{\mathcal{B}\left({\rm R}\right)}$ are the $i$th
known inertial-frame vector, unknown constant bias, and unknown random
noise, respectively, $\forall{\rm v}_{i}^{\mathcal{B}\left({\rm R}\right)},{\rm v}_{i}^{\mathcal{I}\left({\rm R}\right)},{\rm b}_{i}^{\mathcal{B}\left({\rm R}\right)},\omega_{i}^{\mathcal{B}\left({\rm R}\right)}\in\mathbb{R}^{3}$
and $i=1,2,\ldots,N_{{\rm R}}$. Both ${\rm v}_{i}^{\mathcal{I}\left({\rm R}\right)}$
and ${\rm v}_{i}^{\mathcal{B}\left({\rm R}\right)}$ in \eqref{eq:SE3STCH_Vect_R}
can be normalized as
\begin{equation}
\upsilon_{i}^{\mathcal{I}\left({\rm R}\right)}=\frac{{\rm v}_{i}^{\mathcal{I}\left({\rm R}\right)}}{\left\Vert {\rm v}_{i}^{\mathcal{I}\left({\rm R}\right)}\right\Vert },\hspace{1em}\upsilon_{i}^{\mathcal{B}\left({\rm R}\right)}=\frac{{\rm v}_{i}^{\mathcal{B}\left({\rm R}\right)}}{\left\Vert {\rm v}_{i}^{\mathcal{B}\left({\rm R}\right)}\right\Vert }\label{eq:SE3STCH_Vector_norm}
\end{equation}
In that case, $\upsilon_{i}^{\mathcal{I}\left({\rm R}\right)}$ and
$\upsilon_{i}^{\mathcal{B}\left({\rm R}\right)}$ in \eqref{eq:SE3STCH_Vector_norm}
can be utilized to extract the body's attitude instead of ${\rm v}_{i}^{\mathcal{I}\left({\rm R}\right)}$
and ${\rm v}_{i}^{\mathcal{B}\left({\rm R}\right)}$. Define the following
two sets
\begin{equation}
\begin{cases}
\upsilon^{\mathcal{I}\left({\rm R}\right)} & =\left[\upsilon_{1}^{\mathcal{I}\left({\rm R}\right)},\upsilon_{2}^{\mathcal{I}\left({\rm R}\right)},\ldots,\upsilon_{N_{{\rm R}}}^{\mathcal{I}\left({\rm R}\right)}\right]\in\left\{ \mathcal{I}\right\} \\
\upsilon^{\mathcal{B}\left({\rm R}\right)} & =\left[\upsilon_{1}^{\mathcal{B}\left({\rm R}\right)},\upsilon_{2}^{\mathcal{B}\left({\rm R}\right)},\ldots,\upsilon_{N_{{\rm R}}}^{\mathcal{B}\left({\rm R}\right)}\right]\in\left\{ \mathcal{B}\right\} 
\end{cases}\label{eq:SE3STCH_Set_R_Norm}
\end{equation}
where $\upsilon^{\mathcal{I}\left({\rm R}\right)},\upsilon^{\mathcal{B}\left({\rm R}\right)}\in\mathbb{R}^{3\times N_{{\rm R}}}$
contain the normalized vectors introduced in \eqref{eq:SE3STCH_Vector_norm}.
From the other side, the rigid-body's position can be determined if
the body's attitude is available and there are $N_{{\rm L}}$ known
landmarks identified, for instance, by a low-cost inertial vision
system such that the $i$th body-frame measurement is given by
\[
\left[\begin{array}{c}
{\rm v}_{i}^{\mathcal{B}\left({\rm L}\right)}\\
1
\end{array}\right]=\boldsymbol{T}^{-1}\left[\begin{array}{c}
{\rm v}_{i}^{\mathcal{I}\left({\rm L}\right)}\\
1
\end{array}\right]+\left[\begin{array}{c}
{\rm b}_{i}^{\mathcal{B}\left({\rm L}\right)}\\
0
\end{array}\right]+\left[\begin{array}{c}
\omega_{i}^{\mathcal{B}\left({\rm L}\right)}\\
0
\end{array}\right]
\]
or more simply,
\begin{equation}
{\rm v}_{i}^{\mathcal{B}\left({\rm L}\right)}=R^{\top}({\rm v}_{i}^{\mathcal{I}\left({\rm L}\right)}-P)+{\rm b}_{i}^{\mathcal{B}\left({\rm L}\right)}+\omega_{i}^{\mathcal{B}\left({\rm L}\right)}\label{eq:SE3STCH_Vec_Landmark}
\end{equation}
with ${\rm v}_{i}^{\mathcal{I}\left({\rm L}\right)}$ being the $i$th
known landmark placed in the inertial-frame, ${\rm b}_{i}^{\mathcal{B}\left({\rm L}\right)}$
being the additive unknown constant bias, and $\omega_{i}^{\mathcal{B}\left({\rm L}\right)}$
being the additive unknown random noise vector, for all ${\rm v}_{i}^{\mathcal{B}\left({\rm L}\right)},{\rm v}_{i}^{\mathcal{I}\left({\rm L}\right)},{\rm b}_{i}^{\mathcal{B}\left({\rm L}\right)},\omega_{i}^{\mathcal{B}\left({\rm L}\right)}\in\mathbb{R}^{3}$
and $i=1,2,\ldots,N_{{\rm L}}$. The inertial-frame and body-frame
vectors in \eqref{eq:SE3STCH_Vec_Landmark} are divided into the following
two sets

\begin{equation}
\begin{cases}
{\rm v}^{\mathcal{I}\left({\rm L}\right)} & =\left[{\rm v}_{1}^{\mathcal{I}\left({\rm L}\right)},\ldots,{\rm v}_{N_{{\rm L}}}^{\mathcal{I}\left({\rm L}\right)}\right]\in\left\{ \mathcal{I}\right\} \\
{\rm v}^{\mathcal{B}\left({\rm L}\right)} & =\left[{\rm v}_{1}^{\mathcal{B}\left({\rm L}\right)},\ldots,{\rm v}_{N_{{\rm L}}}^{\mathcal{B}\left({\rm L}\right)}\right]\in\left\{ \mathcal{B}\right\} 
\end{cases}\label{eq:SE3STCH_Set_L}
\end{equation}
where ${\rm v}^{\mathcal{I}\left({\rm L}\right)},{\rm v}^{\mathcal{B}\left({\rm L}\right)}\in\mathbb{R}^{3\times N_{{\rm L}}}$.
For the case when more than one landmark is available for measurement,
weighted geometric center approach can be employed
\begin{align}
P_{c}^{\mathcal{I}} & =\frac{1}{\sum_{i=1}^{N_{{\rm L}}}s_{i}^{{\rm L}}}\sum_{i=1}^{N_{{\rm L}}}s_{i}^{{\rm L}}{\rm v}_{i}^{\mathcal{I}\left({\rm L}\right)}\label{eq:SE3STCH_Center_Landmark_I}\\
P_{c}^{\mathcal{B}} & =\frac{1}{\sum_{i=1}^{N_{{\rm L}}}s_{i}^{{\rm L}}}\sum_{i=1}^{N_{{\rm L}}}s_{i}^{{\rm L}}{\rm v}_{i}^{\mathcal{B}\left({\rm L}\right)}\label{eq:SE3STCH_Center_Landmark_B}
\end{align}
where $s_{i}^{{\rm L}}$ refers to the confidence level of the $i$th
measurement.
\begin{assum}
	\label{Assum:SE3STCH_1} The pose of a rigid-body can be obtained
	provided that the set in \eqref{eq:SE3STCH_Set_R_Norm} has rank 3
	and the rank of the set in \eqref{eq:SE3STCH_Set_L} is nonzero such
	that there are at least two non-collinear vectors in \eqref{eq:SE3STCH_Vector_norm}
	($N_{{\rm R}}\geq2$) and one landmark in \eqref{eq:SE3STCH_Vec_Landmark}
	($N_{{\rm L}}\geq1$) available. For $N_{{\rm R}}=2$, the third vector
	can be obtained through $\upsilon_{3}^{\mathcal{I}\left({\rm R}\right)}=\upsilon_{1}^{\mathcal{I}\left({\rm R}\right)}\times\upsilon_{2}^{\mathcal{I}\left({\rm R}\right)}$
	and $\upsilon_{3}^{\mathcal{B}\left({\rm R}\right)}=\upsilon_{1}^{\mathcal{B}\left({\rm R}\right)}\times\upsilon_{2}^{\mathcal{B}\left({\rm R}\right)}$.
\end{assum}
Accordingly, the homogeneous transformation matrix $\boldsymbol{T}$
is obtainable if Assumption \ref{Assum:SE3STCH_1} is valid, (e.g.,
\cite{mohamed2019filters,hashim2019SE3Det,rehbinder2003pose,vasconcelos2010nonlinear,baldwin2009nonlinear,hua2011observer}).
With a view to simplifying the stability analysis, ${\rm v}_{i}^{\mathcal{B}\left({\rm R}\right)}$
and ${\rm v}_{i}^{\mathcal{B}\left({\rm L}\right)}$ are considered
to be noise and bias free. In the Simulation Section, in contrast,
the noise present in the measurements ${\rm v}_{i}^{\mathcal{B}\left({\rm R}\right)}$
and ${\rm v}_{i}^{\mathcal{B}\left({\rm L}\right)}$ is taken into
account. Let us define the pose dynamics with respect to the homogeneous
transformation matrix $\boldsymbol{T}$ \eqref{eq:SE3PPF_T_matrix2}
as
\[
\left[\begin{array}{cc}
\dot{R} & \dot{P}\\
\mathbf{0}_{3\times1}^{\top} & 0
\end{array}\right]=\left[\begin{array}{cc}
R & P\\
\mathbf{0}_{3\times1}^{\top} & 1
\end{array}\right]\left[\begin{array}{cc}
\left[\Omega\right]_{\times} & V\\
\mathbf{0}_{3\times1}^{\top} & 0
\end{array}\right]
\]
with
\begin{align}
\dot{P} & =RV\nonumber \\
\dot{R} & =R\left[\Omega\right]_{\times}\label{eq:SE3PPF_R_Dynamics}\\
\dot{\boldsymbol{T}} & =\boldsymbol{T}\left[\mathcal{Y}\right]_{\wedge}\label{eq:SE3PPF_T_Dynamics}
\end{align}
where $\Omega\in\mathbb{R}^{3}$ represents the true angular velocity,
$V\in\mathbb{R}^{3}$ denotes the translational velocity of the moving
body, and $\mathcal{Y}=\left[\Omega^{\top},V^{\top}\right]^{\top}\in\mathbb{R}^{6}$
denotes the group velocity vector. The measurements of angular and
translational velocities can be expressed, respectively, as
\begin{align}
\Omega_{m} & =\Omega+b_{\Omega}+\omega_{\Omega}\in\left\{ \mathcal{B}\right\} \label{eq:SE3PPF_Angular}\\
V_{m} & =V+b_{V}+\omega_{V}\in\left\{ \mathcal{B}\right\} \label{eq:SE3PPF_V_Trans}
\end{align}
where $b_{\Omega}$ and $b_{V}$ stand for constant bias vectors,
while $\omega_{\Omega}$ and $\omega_{V}$ refer to unknown random
noise attached to the measurement, $\forall b_{\Omega},b_{V},\omega_{\Omega},\omega_{V}\in\mathbb{R}^{3}$.
Define the group vectors of velocity measurements, bias, and noise
as $\mathcal{Y}_{m}=\left[\Omega_{m}^{\top},V_{m}^{\top}\right]^{\top}$,
$b=\left[b_{\Omega}^{\top},b_{V}^{\top}\right]^{\top}$, and $\omega=\left[\omega_{\Omega}^{\top},\omega_{V}^{\top}\right]^{\top}$,
respectively, $\forall\mathcal{Y}_{m},b,\omega\in\mathbb{R}^{6}$.
$\omega$ being a random Gaussian noise vector has zero mean and is
bounded. Since the derivative of a Gaussian process results in a Gaussian
process \cite{khasminskii1980stochastic,jazwinski2007stochastic},
one could define $\omega$ as a function of a Brownian motion process
vector such that
\begin{align}
\omega & =\mathcal{Q}\frac{d\beta}{dt},\hspace{1em}\text{with }\omega_{\Omega}=\mathcal{Q}_{\Omega}\frac{d\beta_{\Omega}}{dt}\text{ and }\omega_{V}=\mathcal{Q}_{V}\frac{d\beta_{V}}{dt}\label{eq:SE3PPF_noise}
\end{align}
where $\beta=\left[\beta_{\Omega}^{\top},\beta_{V}^{\top}\right]^{\top}\in\mathbb{R}^{6}$,
and $\mathcal{Q}=\left[\begin{array}{cc}
\mathcal{Q}_{\Omega} & \mathbf{0}_{3\times3}\\
\mathbf{0}_{3\times3} & \mathcal{Q}_{V}
\end{array}\right]\in\mathbb{R}^{6\times6}$ is a diagonal matrix whose diagonal includes unknown time-variant
non-negative components for all $\beta_{\Omega},\beta_{V}\in\mathbb{R}^{3}$
and $\mathcal{Q}_{\Omega},\mathcal{Q}_{V}\in\mathbb{R}^{3\times3}$.
Brownian motion process signal is characterized by the following properties
\cite{jazwinski2007stochastic,ito1984lectures,deng2001stabilization}
\[
\mathbb{P}\left\{ \beta\left(0\right)=0\right\} =1,\hspace{1em}\mathbb{E}\left[d\beta/dt\right]=0,\hspace{1em}\mathbb{E}\left[\beta\right]=0
\]
In the light of the identity in \eqref{eq:SE3STCH_Identity6}, the
expression of $\left\Vert R\right\Vert _{{\rm I}}$ in \eqref{eq:SE3STCH_Ecul_Dist},
and the expressions in \eqref{eq:SE3PPF_Angular} and \eqref{eq:SE3PPF_noise},
the true attitude dynamics in \eqref{eq:SE3PPF_R_Dynamics} can be
written in terms of \eqref{eq:SE3STCH_Ecul_Dist} in incremental form
as
\begin{align}
d||R||_{{\rm I}} & =-\frac{1}{4}{\rm Tr}\left\{ dR\right\} \nonumber \\
& =-\frac{1}{4}{\rm Tr}\left\{ \boldsymbol{\mathcal{P}}_{a}\left(R\right)\left[\Omega\right]_{\times}\right\} dt\nonumber \\
& =\frac{1}{2}\mathbf{vex}\left(\boldsymbol{\mathcal{P}}_{a}\left(R\right)\right)^{\top}\left(\left(\Omega_{m}-b_{\Omega}\right)dt-\mathcal{Q}_{\Omega}d\beta_{\Omega}\right)\label{eq:SE3PPF_NormR_dynam}
\end{align}
Define $\mathcal{X}=\left[||R||_{{\rm I}},P^{\top}\right]^{\top}\in\mathbb{R}^{4}$.
Thus, from \eqref{eq:SE3PPF_NormR_dynam}, the pose dynamics in \eqref{eq:SE3PPF_T_Dynamics}
are written in vector form as a stochastic differential equation%
\begin{align}
d\mathcal{X}= & \mathcal{F}dt-\mathcal{G}\mathcal{Q}d\beta\nonumber \\
\mathcal{G}= & \left[\begin{array}{cc}
\frac{1}{2}\boldsymbol{\Upsilon}_{a}^{\top}\left(R\right) & \mathbf{0}_{3\times1}^{\top}\\
\mathbf{0}_{3\times3} & R
\end{array}\right]\nonumber \\
\mathcal{F}= & \mathcal{G}(\mathcal{Y}_{m}-b)\label{eq:SE3PPF_T_VEC_Dyn}
\end{align}
where both $\mathcal{G}$ and $\mathcal{F}$ are locally Lipschitz.

\begin{rem}
	\label{rem:Unstable-set}Define $\mathcal{S}_{0}\subseteq\mathbb{SO}\left(3\right)\times\mathbb{R}^{3}$
	as a non-attractive, forward invariant unstable set:
	\begin{equation}
	\mathcal{S}_{0}=\{\left.\left(R\left(0\right),P\left(0\right)\right)\right|{\rm Tr}\left\{ R\left(0\right)\right\} =-1,P\left(0\right)=\mathbf{0}_{3\times1}\}\label{eq:SO3_PPF_STCH_SET}
	\end{equation}
	where the only three possible scenarios for ${\rm Tr}\left\{ R\left(0\right)\right\} =-1$
	are: $R\left(0\right)={\rm diag}(1,-1,-1)$, $R\left(0\right)={\rm diag}(-1,1,-1)$,
	and $R\left(0\right)={\rm diag}(-1,-1,1)$.
\end{rem}
The stochastic differential equation of the system in \eqref{eq:SE3PPF_T_VEC_Dyn}
has a solution on $t\in\left[t\left(0\right),T\right]\forall t\left(0\right)\leq T<\infty$
and $R\left(0\right)\notin\mathcal{S}_{0}$ in the mean square sense.
Additionally, for any $\mathcal{X}\left(t\right)$ where $t\neq t\left(0\right)$,
$\mathcal{X}-\mathcal{X}\left(0\right)$ is independent of $\beta\left(\tau\right)$
$\forall\tau\geq t$ and $\forall t\in\left[t\left(0\right),T\right]$
(Theorem 4.5 \cite{jazwinski2007stochastic}). The goal of this work
is to design a reliable pose estimator that achieves adaptive stabilization
and accounts for unknown constant bias and unknown time-variant covariance
matrix attached to velocity measurements. Let the upper-bound of the
diagonal entries in $\mathcal{Q}_{\Omega}^{2}$ and $\mathcal{Q}_{V}^{2}$
be $\sigma$ and $\xi$, respectively, with $\sigma,\xi\in\mathbb{R}^{3}$
such that{\small{}
	\begin{align}
	\sigma & =\left[{\rm max}\{\mathcal{Q}_{\Omega\left(1,1\right)}^{2}\},{\rm max}\{\mathcal{Q}_{\Omega\left(2,2\right)}^{2}\},{\rm max}\{\mathcal{Q}_{\Omega\left(3,3\right)}^{2}\}\right]^{\top}\label{eq:SE3STCH_s_max}\\
	\xi & =\left[{\rm max}\{\mathcal{Q}_{V\left(1,1\right)}^{2}\},{\rm max}\{\mathcal{Q}_{V\left(2,2\right)}^{2}\},{\rm max}\{\mathcal{Q}_{V\left(3,3\right)}^{2}\}\right]^{\top}\label{eq:SE3STCH_beta_max}
	\end{align}
}with ${\rm max}\left\{ \cdot\right\} $ being the maximum value of
the element.
\begin{assum}
	\label{Assum:SE3STCH_2} Consider $b$, $\sigma$, and $\xi$ to be
	upper-bounded by $\Gamma$ and to belong to a compact set $\Delta$
	with $\Gamma\in\mathbb{R}_{+}$ and $\left\Vert \Delta\right\Vert \leq\Gamma<\infty$.
\end{assum}
\begin{defn}
	\label{def:SE3STCH_1}\cite{ji2006adaptive} For $\mathcal{X}=\left[||R||_{I},P^{\top}\right]^{\top}$
	in the stochastic differential system \eqref{eq:SE3PPF_T_VEC_Dyn},
	define a compact set $\Theta\in\mathbb{R}^{4}$ and $\mathcal{X}\left(0\right)=\mathcal{X}\left(t\left(0\right)\right)$.
	If there exists a positive constant $c$ and a time constant $t_{c}=t_{c}\left(c,\mathcal{X}\left(0\right)\right)$
	with $\mathbb{E}\left[\left\Vert \mathcal{X}\right\Vert \right]<c,\forall t>t\left(0\right)+t_{c}$,
	the trajectory of $\mathcal{X}$ is semi-globally uniformly ultimately
	bounded (SGUUB). 
\end{defn}
\begin{defn}
	\label{def:SE3STCH_2} Consider the stochastic dynamics in \eqref{eq:SE3PPF_T_VEC_Dyn}
	and let $V\left(\mathcal{X}\right)$ be a given function which is
	twice differentiable such that $V\left(\mathcal{X}\right)\in\mathcal{C}^{2}$.
	The differential operator of $V\left(\mathcal{X}\right)$ is defined
	by
	\[
	\mathcal{L}V\left(\mathcal{X}\right)=V_{\mathcal{X}}^{\top}\mathcal{F}+\frac{1}{2}{\rm Tr}\left\{ \mathcal{G}\mathcal{Q}^{2}\mathcal{G}^{\top}V_{\mathcal{X}\mathcal{X}}\right\} 
	\]
	where $V_{\mathcal{X}}=\partial V/\partial\mathcal{X}$ and $V_{\mathcal{X}}=\partial^{2}V/\partial\mathcal{X}^{2}$. 
\end{defn}
\begin{lem}
	\label{lem:SE3STCH_1} \cite{deng2001stabilization,ji2006adaptive,deng1997stochastic}
	Consider the stochastic dynamics in \eqref{eq:SE3PPF_T_VEC_Dyn} and
	suppose that there exists a potential function $V\left(\mathcal{X}\right)$
	that satisfies $V\in\mathcal{C}^{2}$ with $V:\mathbb{R}^{4}\rightarrow\mathbb{R}_{+}$.
	Suppose there are a class $\mathcal{K}_{\infty}$ function $\bar{\upsilon}_{1}\left(\cdot\right)$
	and $\bar{\upsilon}_{2}\left(\cdot\right)$, constants $\boldsymbol{{\rm c}}>0$
	and $\mathbf{k}\geq0$, and a non-negative function $\boldsymbol{\mathcal{N}}\left(\left\Vert \mathcal{X}\right\Vert \right)$
	such that
	\begin{equation}
	\bar{\upsilon}_{1}\left(\left\Vert \mathcal{X}\right\Vert \right)\leq V\leq\bar{\upsilon}_{2}\left(\left\Vert \mathcal{X}\right\Vert \right)\label{eq:SE3STCH_Vfunction_Lyap}
	\end{equation}
	\begin{align}
	\mathcal{L}V\left(\mathcal{X}\right)= & V_{\mathcal{X}}^{\top}\mathcal{F}\left(\mathcal{X}\right)+\frac{1}{2}{\rm Tr}\left\{ \mathcal{G}\mathcal{Q}^{2}\mathcal{G}^{\top}V_{\mathcal{X}\mathcal{X}}\right\} \nonumber \\
	\leq & -\boldsymbol{{\rm c}}\boldsymbol{\mathcal{N}}\left(\left\Vert \mathcal{X}\right\Vert \right)+\mathbf{k}\label{eq:SE3STCH_dVfunction_Lyap}
	\end{align}
	Then for $\mathcal{X}\left(0\right)\in\mathbb{R}^{4}$ and $R\left(0\right)\notin\mathcal{S}_{0}$
	defined in Remark \ref{rem:Unstable-set}, there exists almost a unique
	strong solution on $\left[0,\infty\right)$ for the dynamic system
	in \eqref{eq:SE3PPF_T_VEC_Dyn}. Also, the solution $\mathcal{X}$
	of the stochastic system in \eqref{eq:SE3PPF_T_VEC_Dyn} is bounded
	in probability satisfying
	\begin{equation}
	\mathbb{E}\left[V\left(\mathcal{X}\right)\right]\leq V\left(\mathcal{X}\left(0\right)\right){\rm exp}\left(-\boldsymbol{{\rm c}}t\right)+\frac{\mathbf{k}}{\boldsymbol{{\rm c}}}\label{eq:SE3STCH_EVfunction_Lyap}
	\end{equation}
	with $\mathcal{X}\in\mathbb{R}^{4}$ being SGUUB%
	.
\end{lem}
The proof of Lemma \ref{lem:SE3STCH_1} can be found in \cite{deng2001stabilization}.
For $R\in\mathbb{SO}\left(3\right)$, the set $\mathcal{S}_{0}$ is
unstable and forward invariant for the stochastic system described
in \eqref{eq:SE3PPF_T_Dynamics} and \eqref{eq:SE3PPF_T_VEC_Dyn}
\cite{shuster1993survey}. From almost any initial condition given
that $R\left(0\right)\notin\mathcal{S}_{0}$, we have $-1<{\rm Tr}\left\{ R\left(0\right)\right\} \leq3$
and the trajectory of $\mathcal{X}$ is SGUUB%
.
\begin{lem}
	\label{lem:SE3STCH_2} (Young's inequality) Suppose there are two
	real vectors $x$ and $y$ with $x,y\in\mathbb{R}^{n}$. For any $a>0$
	and $b>0$ that satisfy $\frac{1}{a}+\frac{1}{b}=1$, there is
	\begin{equation}
	x^{\top}y\leq\left(1/a\right)\varrho^{a}\left\Vert x\right\Vert ^{a}+\left(1/b\right)\varrho^{-b}\left\Vert y\right\Vert ^{b}\label{eq:SE3STCH_lem_ineq}
	\end{equation}
	where $\varrho$ is a small positive constant.
\end{lem}
\begin{lem}
	\label{Lemm:SE3STCH_1}Consider $R\in\mathbb{SO}\left(3\right)$,
	$\mathbf{M}_{{\rm R}}=\mathbf{M}_{{\rm R}}^{\top}\in\mathbb{R}^{3\times3}$
	with a rank of $3$, ${\rm Tr}\{\mathbf{M}_{{\rm R}}\}=3$, and $\bar{\mathbf{M}}_{{\rm R}}={\rm Tr}\{\mathbf{M}_{{\rm R}}\}\mathbf{I}_{3}-\mathbf{M}_{{\rm R}}$
	with the minimum singular value of $\bar{\mathbf{M}}_{{\rm R}}$ being
	$\underline{\lambda}_{1}=\underline{\lambda}(\bar{\mathbf{M}}_{{\rm R}})$.
	Then, the following holds:
	\begin{align}
	||\mathbf{vex}(\boldsymbol{\mathcal{P}}_{a}\left(R\right))||^{2} & =4\left(1-||R||_{{\rm I}}\right)||R||_{{\rm I}}\label{eq:SE3PPF_lemm1_1}\\
	\frac{2}{\underline{\lambda}_{1}}\frac{||\mathbf{vex}(\boldsymbol{\mathcal{P}}_{a}(R\mathbf{M}_{{\rm R}}))||^{2}}{1+{\rm Tr}\{R\mathbf{M}_{{\rm R}}\mathbf{M}_{{\rm R}}^{-1}\}} & \geq\left\Vert R\mathbf{M}_{{\rm R}}\right\Vert _{{\rm I}}\label{eq:SE3PPF_lemm1_2}
	\end{align}
	\textbf{Proof. See \nameref{sec:SO3_PPF_STCH_AppendixA}.} 
\end{lem}

\section{Nonlinear Stochastic Pose Estimators on $\mathbb{SE}\left(3\right)$
	\label{sec:SE3_Stochastic-Complementary-Filters}}

This section presents two nonlinear stochastic pose estimators evolved
directly on $\mathbb{SE}\left(3\right)$ designed with reliability
as the primary consideration. The first estimator is termed a semi-direct
pose estimator since it requires the attitude and position to be reestablished
using vector measurements in \eqref{eq:SE3STCH_Set_R_Norm} and \eqref{eq:SE3STCH_Set_L}
and the group velocity measurements described in \eqref{eq:SE3PPF_Angular}
and \eqref{eq:SE3PPF_V_Trans}. Whereas, the second pose estimator
is referred to as direct and is designed to use the above-mentioned
measurements directly. Define the estimate of the homogeneous transformation
matrix by
\[
\hat{\boldsymbol{T}}=\left[\begin{array}{cc}
\hat{R} & \hat{P}\\
\mathbf{0}_{3\times1}^{\top} & 1
\end{array}\right]\in\mathbb{SE}\left(3\right)
\]
The proposed pose estimators are evolved on $\mathbb{SE}\left(3\right)$
and their structure follows
\[
\dot{\hat{\boldsymbol{T}}}=\hat{\boldsymbol{T}}\left[\hat{\mathcal{Y}}\right]_{\wedge}
\]
where $\hat{\mathcal{Y}}=[\hat{\Omega}^{\top},\hat{V}^{\top}]^{\top}\in\mathbb{R}^{6}$
such that $\dot{\hat{R}}=\hat{R}[\hat{\Omega}]_{\times}$ and $\dot{\hat{P}}=\hat{R}\hat{V}$.
Consider the error of the homogeneous transformation matrix estimation
to be given by
\begin{align}
\tilde{\boldsymbol{T}} & =\hat{\boldsymbol{T}}\boldsymbol{T}^{-1}=\left[\begin{array}{cc}
\tilde{R} & \tilde{P}\\
\mathbf{0}_{3\times1}^{\top} & 1
\end{array}\right]\label{eq:SE3STCH_T_error}
\end{align}
where $\tilde{R}=\hat{R}R^{\top}$ and $\tilde{P}=\hat{P}-\tilde{R}P$
are the orientation and the position error, respectively, between
the rigid-body-frame and the estimator-frame. As such, driving $\hat{\boldsymbol{T}}\rightarrow\boldsymbol{T}$
ensures that $\tilde{P}\rightarrow\mathbf{0}_{3\times1}$ and $\tilde{R}\rightarrow\mathbf{I}_{3}$,
or equivalently, $||\tilde{R}||_{I}=\frac{1}{4}{\rm Tr}\{\mathbf{I}_{3}-\tilde{R}\}\rightarrow0$,
which implies driving $\tilde{\boldsymbol{T}}\rightarrow\mathbf{I}_{4}$.
Consider the estimates of the unknown parameters $b$ and $\sigma$
to be denoted, respectively, by $\hat{b}=\left[\hat{b}_{\Omega}^{\top},\hat{b}_{V}^{\top}\right]^{\top}$
and $\hat{\sigma}$ for all $\hat{b}_{\Omega},\hat{b}_{V},\hat{\sigma}\in\mathbb{R}^{3}$.
Consider the error in $b$ and $\sigma$ to be
\begin{align}
\tilde{b} & =b-\hat{b}\label{eq:SE3STCH_bt}\\
\tilde{\sigma} & =\sigma-\hat{\sigma}\label{eq:SE3STCH_st}
\end{align}
where $\tilde{b}=\left[\tilde{b}_{\Omega}^{\top},\tilde{b}_{V}^{\top}\right]^{\top}$
for all $\tilde{b}_{\Omega},\tilde{b}_{V},\tilde{\sigma}\in\mathbb{R}^{3}$.

\subsection{Semi-direct Nonlinear Stochastic Pose Estimator on $\mathbb{SE}\left(3\right)$
	\label{subsec:SE3_Passive-Filter}}

Let the reconstructed matrix of the true homogeneous transformation
matrix be denoted by $\boldsymbol{T}_{y}=\left[\begin{array}{cc}
R_{y} & P_{y}\\
\mathbf{0}_{3\times1}^{\top} & 1
\end{array}\right]$. In this context, $R_{y}$ refers to uncertain attitude which can
be reconstructed, for instance \cite{shuster1981three,markley1988attitude} and for attitude construction methods visit \cite{hashim2020AtiitudeSurvey}.
From \eqref{eq:SE3STCH_Center_Landmark_I} and \eqref{eq:SE3STCH_Center_Landmark_B},
$P_{y}$ can be reconstructed using $P_{y}=\frac{1}{\sum_{i=1}^{N_{{\rm L}}}k_{i}^{{\rm L}}}\sum_{i=1}^{N_{{\rm L}}}s_{i}^{{\rm L}}\left({\rm v}_{i}^{\mathcal{I}\left({\rm L}\right)}-R_{y}{\rm v}_{i}^{\mathcal{B}\left({\rm L}\right)}\right)$.
From \eqref{eq:SE3STCH_T_error} and in view of the pose dynamics
in \eqref{eq:SE3PPF_T_VEC_Dyn}, one can rewrite the error in vector
form as
\begin{equation}
\mathcal{E}=\left[\mathcal{E}_{R},\mathcal{E}_{P}^{\top}\right]^{\top}=\left[||\tilde{R}||_{{\rm I}},\tilde{P}^{\top}\right]^{\top}\in\mathbb{R}^{4}\label{eq:SE3STCH_E1}
\end{equation}
where $\tilde{R}=\hat{R}R_{y}^{\top}$, $\mathcal{E}_{R}=||\tilde{R}||_{{\rm I}}=\frac{1}{4}{\rm Tr}\left\{ \mathbf{I}_{3}-\tilde{R}\right\} $
as defined in \eqref{eq:SE3STCH_Ecul_Dist}, and $\mathcal{E}_{P}=\tilde{P}=\hat{P}-\tilde{R}P_{y}$.
Consider the following nonlinear pose estimator on $\mathbb{SE}\left(3\right)$
\begin{equation}
\left[\begin{array}{cc}
\dot{\hat{R}} & \dot{\hat{P}}\\
\mathbf{0}_{3\times1}^{\top} & 0
\end{array}\right]=\left[\begin{array}{cc}
\hat{R} & \hat{P}\\
\mathbf{0}_{3\times1}^{\top} & 1
\end{array}\right]\left[\begin{array}{c}
\Omega_{m}-\hat{b}_{\Omega}-W_{\Omega}\\
V_{m}-\hat{b}_{V}-W_{V}
\end{array}\right]_{\land}\label{eq:SE3PPF_Ty_dot}
\end{equation}
\begin{align}
W_{\Omega}= & \frac{2k_{w}}{1-\mathcal{E}_{R}}\left[\hat{R}^{\top}\boldsymbol{\Upsilon}_{a}\left(\tilde{R}\right)\right]_{{\rm D}}\hat{\sigma}\label{eq:SE3STCH_Wy_om}\\
W_{V}= & -\hat{R}^{\top}\left[\hat{P}\right]_{\times}\hat{R}W_{\Omega}+\frac{k_{w}}{\varrho}\hat{R}^{\top}\mathcal{E}_{P}\label{eq:SE3STCH_Wy_v}\\
\dot{\hat{b}}_{\Omega}= & \frac{\gamma_{b}}{2}(1+\mathcal{E}_{R})\exp(\mathcal{E}_{R})\hat{R}^{\top}\boldsymbol{\Upsilon}_{a}(\tilde{R})\nonumber \\
& -\gamma_{b}\left\Vert \mathcal{E}_{P}\right\Vert ^{2}\hat{R}^{\top}\left[\hat{P}\right]_{\times}\mathcal{E}_{P}-\gamma_{b}k_{b}\hat{b}_{\Omega}\label{eq:SE3STCH_by_om}\\
\dot{\hat{b}}_{V}= & \gamma_{b}\left\Vert \mathcal{E}_{P}\right\Vert ^{2}\hat{R}^{\top}\mathcal{E}_{P}-\gamma_{b}k_{b}\hat{b}_{V}\label{eq:SE3STCH_by_v}\\
K_{\mathcal{E}}= & \gamma_{\sigma}\frac{1+\mathcal{E}_{R}}{1-\mathcal{E}_{R}}\exp\left(\mathcal{E}_{R}\right)\nonumber \\
\dot{\hat{\sigma}}= & k_{w}K_{\mathcal{E}}\left[\hat{R}^{\top}\boldsymbol{\Upsilon}_{a}(\tilde{R})\right]_{{\rm D}}\hat{R}^{\top}\boldsymbol{\Upsilon}_{a}(\tilde{R})-\gamma_{\sigma}k_{\sigma}\hat{\sigma}\label{eq:SE3STCH_sy_om}
\end{align}
where $\mathcal{E}$ is given in \eqref{eq:SE3STCH_E1}, $\boldsymbol{\Upsilon}_{a}(\tilde{R})=\mathbf{vex}(\boldsymbol{\mathcal{P}}_{a}(\tilde{R}))$
is defined in \eqref{eq:SE3STCH_VEX_a}, and $\left[\cdot\right]_{{\rm D}}$
is a diagonal matrix of a vector. $k_{w}$, $\gamma_{b}$, and $\gamma_{\sigma}$
are positive constants, $\hat{b}=\left[\hat{b}_{\Omega}^{\top},\hat{b}_{V}^{\top}\right]^{\top}$
is the estimate of $b$ and $\sigma$ is the estimate of $\hat{\sigma}$. The equivalent quaternion representation and complete implementation
steps of the semi-direct filter are given in \nameref{sec:SO3_PPF_STCH_AppendixB}.
\begin{thm}
	\label{thm:SE3STCH_1}\textbf{ }Consider the pose dynamics in \eqref{eq:SE3PPF_T_Dynamics}
	combined with the group velocity measurements $\mathcal{Y}_{m}=\left[\Omega_{m}^{\top},V_{m}^{\top}\right]^{\top}$
	in \eqref{eq:SE3PPF_Angular} and \eqref{eq:SE3PPF_V_Trans}. Let
	Assumption \ref{Assum:SE3STCH_1} hold. Suppose that $\boldsymbol{T}_{y}$
	is reconstructed based on the vector measurements in \eqref{eq:SE3STCH_Vec_Landmark}
	and \eqref{eq:SE3STCH_Vector_norm}, and geared with the estimator
	in \eqref{eq:SE3PPF_Ty_dot}, \eqref{eq:SE3STCH_Wy_om}, \eqref{eq:SE3STCH_Wy_v},
	\eqref{eq:SE3STCH_by_om}, \eqref{eq:SE3STCH_by_v}, and \eqref{eq:SE3STCH_sy_om}.
	Suppose that the design parameters are selected as follows: $\gamma_{b}>0$,
	$\gamma_{\sigma}>0$, $k_{b}>0$, $k_{\sigma}>0$, $\varrho>0$, and
	$k_{w}>9/8$ with $\varrho$ being selected sufficiently small, and
	recall the set in Remark \ref{rem:Unstable-set}. In case where $\mathcal{Y}_{m}$
	is biased and contaminated by random Gaussian noise $\left(\omega\neq0\right)$,
	and $\tilde{R}\left(0\right)\notin\mathcal{S}_{0}$, all the closed-loop
	signals are semi-globally uniformly ultimately bounded in mean square.
	Additionally, the filter errors could be minimized by the appropriated
	selection of the design parameters.
\end{thm}
\textbf{Proof. }Recall the true and the estimated attitude dynamics
in \eqref{eq:SE3PPF_R_Dynamics} and \eqref{eq:SE3PPF_Ty_dot}, respectively.
Considering that $\tilde{R}=R\hat{R}^{\top}$, the error in attitude
dynamics is
\begin{align}
d\tilde{R} & =d\hat{R}R^{\top}+\hat{R}dR^{\top}\nonumber \\
& =\hat{R}\left[\tilde{b}_{\Omega}-W_{\Omega}\right]_{\times}R^{\top}dt+\hat{R}\left[\mathcal{Q}_{\Omega}d\beta_{\Omega}\right]_{\times}R^{\top}\nonumber \\
& =\left[\hat{R}(\tilde{b}_{\Omega}-W_{\Omega})\right]_{\times}\tilde{R}dt+\left[\hat{R}\mathcal{Q}_{\Omega}d\beta_{\Omega}\right]_{\times}\tilde{R}\label{eq:SE3STCH_Rt_dot}
\end{align}
In the light of \eqref{eq:SE3PPF_R_Dynamics} and \eqref{eq:SE3PPF_NormR_dynam},
and with the aid of the identity in \eqref{eq:SE3STCH_Identity6},
the error dynamics in \eqref{eq:SE3STCH_Rt_dot} can be expressed
in terms of normalized Euclidean distance
\begin{align}
d\left\Vert \tilde{R}\right\Vert _{{\rm I}}= & -\frac{1}{4}{\rm Tr}\left\{ \left[\hat{R}(\tilde{b}_{\Omega}-W_{\Omega})dt+\hat{R}\mathcal{Q}_{\Omega}d\beta_{\Omega}\right]_{\times}\tilde{R}\right\} \nonumber \\
= & \frac{1}{2}\boldsymbol{\Upsilon}_{a}^{\top}(\tilde{R})\hat{R}\left((\tilde{b}_{\Omega}-W_{\Omega})dt+\mathcal{Q}_{\Omega}d\beta_{\Omega}\right)\label{eq:SE3STCH_RtI2_dot}
\end{align}
Given that $\tilde{P}=P-\tilde{R}\hat{P}$, the position dynamics
error can be found in the following way
\begin{align}
d\tilde{P}= & d\hat{P}-d\tilde{R}P-\tilde{R}dP\nonumber \\
= & \hat{R}(\tilde{b}_{V}-W_{V})dt+\left[\hat{R}(\tilde{b}_{\Omega}-W_{\Omega})dt+\hat{R}\mathcal{Q}_{\Omega}d\beta_{\Omega}\right]_{\times}\tilde{P}\nonumber \\
& +\hat{R}\mathcal{Q}_{V}d\beta_{V}-\left[\hat{R}(\tilde{b}_{\Omega}-W_{\Omega})dt+\hat{R}\mathcal{Q}_{\Omega}d\beta_{\Omega}\right]_{\times}\hat{P}\nonumber \\
= & \left(\hat{R}(\tilde{b}_{V}-W_{V})+[\hat{P}-\tilde{P}]_{\times}\hat{R}(\tilde{b}_{\Omega}-W_{\Omega})\right)dt\nonumber \\
& +[\hat{P}-\tilde{P}]_{\times}\hat{R}\mathcal{Q}_{\Omega}d\beta_{\Omega}+\hat{R}\mathcal{Q}_{V}d\beta_{V}\label{eq:SE3STCH_Pt_dot}
\end{align}
Defining $\mathcal{E}=\left[\mathcal{E}_{R},\mathcal{E}_{P}^{\top}\right]^{\top}=\left[||\tilde{R}||_{{\rm I}},\tilde{P}^{\top}\right]^{\top}$
as in \eqref{eq:SE3STCH_E1} and combining it with \eqref{eq:SE3PPF_T_VEC_Dyn},
the following set of equations is obtained
\begin{align}
d\mathcal{E}= & \tilde{\mathcal{F}}dt+\tilde{\mathcal{G}}\mathcal{Q}d\beta\label{eq:SE3STCH_E_dot}\\
\tilde{\mathcal{G}}= & \left[\begin{array}{cc}
\frac{1}{2}\boldsymbol{\Upsilon}_{a}^{\top}(\tilde{R})\hat{R} & \mathbf{0}_{1\times3}\\{}
[\hat{P}-\tilde{P}]_{\times}\hat{R} & \hat{R}
\end{array}\right]\nonumber \\
\tilde{\mathcal{F}}= & \tilde{\mathcal{G}}((\tilde{b}-W)dt+\mathcal{Q}d\beta)\nonumber 
\end{align}
For $V:=V(\mathcal{E},\tilde{b},\tilde{\sigma})$, consider the following
Lyapunov candidate function
\begin{align}
V= & \mathcal{E}_{R}\exp(\mathcal{E}_{R})+\frac{1}{4}\left\Vert \mathcal{E}_{P}\right\Vert ^{4}+\frac{1}{2\gamma_{b}}\left\Vert \tilde{b}\right\Vert ^{2}+\frac{1}{2\gamma_{\sigma}}\left\Vert \tilde{\sigma}\right\Vert ^{2}\label{eq:SE3STCH_V}
\end{align}
The differential operator $\mathcal{L}V$ in Definition \ref{def:SE3STCH_2}
becomes
\begin{align}
\mathcal{L}V= & V_{\mathcal{E}}^{\top}\tilde{\mathcal{F}}+\frac{1}{2}{\rm Tr}\left\{ \tilde{\mathcal{G}}\mathcal{Q}^{2}\tilde{\mathcal{G}}^{\top}V_{\mathcal{E}\mathcal{E}}\right\} -\frac{1}{\gamma_{1}}\tilde{b}^{\top}\dot{\hat{b}}-\frac{1}{\gamma_{2}}\tilde{\sigma}^{\top}\dot{\hat{\sigma}}\label{eq:SE3STCH_VL1}
\end{align}
where $V_{\mathcal{E}}=\partial V/\partial\mathcal{E}$ and $V_{\mathcal{E}\mathcal{E}}=\partial^{2}V/\partial\mathcal{E}^{2}$.
It could be easily shown that the first and second partial derivatives
of \eqref{eq:SE3STCH_V} can be expressed with respect to $\mathcal{E}$
as shown below 
\begin{align}
V_{\mathcal{E}}= & \left[\begin{array}{cc}
1+\mathcal{E}_{R} & \mathbf{0}_{3\times1}^{\top}\\
\mathbf{0}_{3\times1} & ||\mathcal{E}_{P}||^{2}\mathbf{I}_{3}
\end{array}\right]\left[\begin{array}{c}
\exp(\mathcal{E}_{R})\\
\mathcal{E}_{P}
\end{array}\right]\label{eq:SE3STCH_Vv}\\
V_{\mathcal{E}\mathcal{E}}= & \left[\begin{array}{cc}
(2+\mathcal{E}_{R})\exp(\mathcal{E}_{R}) & \mathbf{0}_{3\times1}^{\top}\\
\mathbf{0}_{3\times1} & ||\mathcal{E}_{P}||^{2}\mathbf{I}_{3}+2\mathcal{E}_{P}\mathcal{E}_{P}^{\top}
\end{array}\right]\label{eq:SE3STCH_Vvv}
\end{align}
Thus, using \eqref{eq:SE3STCH_Vv} and \eqref{eq:SE3STCH_Vvv}, the
differential operator $\mathcal{L}V$ in \eqref{eq:SE3STCH_VL1} can
be rewritten as
\begin{align}
\mathcal{L}V= & \frac{1}{2}(1+\mathcal{E}_{R})\exp(\mathcal{E}_{R})\boldsymbol{\Upsilon}_{a}^{\top}(\tilde{R})\hat{R}(\tilde{b}_{\Omega}-W_{\Omega})\nonumber \\
& +\frac{3}{8}(2+\mathcal{E}_{R})\exp(\mathcal{E}_{R})\boldsymbol{\Upsilon}_{a}^{\top}(\tilde{R})\hat{R}\mathcal{Q}_{\Omega}^{2}\hat{R}^{\top}\boldsymbol{\Upsilon}_{a}(\tilde{R})\nonumber \\
& +||\mathcal{E}_{P}||^{2}\mathcal{E}_{P}^{\top}[\hat{P}-\tilde{P}]_{\times}\hat{R}(\tilde{b}_{\Omega}-W_{\Omega})\nonumber \\
& +||\mathcal{E}_{P}||^{2}\mathcal{E}_{P}^{\top}\hat{R}(\tilde{b}_{V}-W_{V})\nonumber \\
& +\frac{1}{2}{\rm Tr}\left\{ \left(\left\Vert \mathcal{E}_{P}\right\Vert ^{2}\mathbf{I}_{3}+2\mathcal{E}_{P}\mathcal{E}_{P}^{\top}\right)\hat{R}\mathcal{Q}_{V}^{2}\hat{R}^{\top}\right\} \nonumber \\
& +\frac{1}{2}{\rm Tr}\left\{ (||\mathcal{E}_{P}||^{2}\mathbf{I}_{3}+2\mathcal{E}_{P}\mathcal{E}_{P}^{\top})[\hat{P}-\tilde{P}]_{\times}\hat{R}\mathcal{Q}_{V}^{2}\hat{R}^{\top}\right.\nonumber \\
& \hspace{5em}\times\left.[\hat{P}-\tilde{P}]_{\times}^{\top}\right\} -\frac{1}{\gamma_{b}}\tilde{b}^{\top}\dot{\hat{b}}-\frac{1}{\gamma_{\sigma}}\tilde{\sigma}^{\top}\dot{\hat{\sigma}}\label{eq:SE3STCH_VL2}
\end{align}
Since $\hat{R}\mathcal{Q}_{V}^{2}\hat{R}^{\top}$ is positive semi-definite,
the last trace component in \eqref{eq:SE3STCH_VL2} is negative semi-definite.
Also, in the light of the fact that $\tilde{P}^{\top}[\tilde{P}]_{\times}=\mathbf{0}_{1\times3}$,
the differential operator in \eqref{eq:SE3STCH_VL2} can take a form
of an inequality
\begin{align}
\mathcal{L}V\leq & \frac{1}{2}(1+\mathcal{E}_{R})\exp(\mathcal{E}_{R})\boldsymbol{\Upsilon}_{a}^{\top}\left(\tilde{R}\right)\hat{R}(\tilde{b}_{\Omega}-W_{\Omega})\nonumber \\
& +\frac{3}{8}(2+\mathcal{E}_{R})\exp(\mathcal{E}_{R})\boldsymbol{\Upsilon}_{a}^{\top}(\tilde{R})\hat{R}\left[\sigma\right]_{{\rm D}}\hat{R}^{\top}\boldsymbol{\Upsilon}_{a}\left(\tilde{R}\right)\nonumber \\
& +\left\Vert \mathcal{E}_{P}\right\Vert ^{2}\mathcal{E}_{P}^{\top}\left([\hat{P}]_{\times}\hat{R}(\tilde{b}_{\Omega}-W_{\Omega})+\hat{R}(\tilde{b}_{V}-W_{V})\right)\nonumber \\
& +\frac{1}{2}{\rm Tr}\left\{ \left(\left\Vert \mathcal{E}_{P}\right\Vert ^{2}\mathbf{I}_{3}+2\mathcal{E}_{P}\mathcal{E}_{P}^{\top}\right)\hat{R}\left[\xi\right]_{{\rm D}}\hat{R}^{\top}\right\} \nonumber \\
& -\frac{1}{\gamma_{b}}\tilde{b}^{\top}\dot{\hat{b}}-\frac{1}{\gamma_{\sigma}}\tilde{\sigma}^{\top}\dot{\hat{\sigma}}\label{eq:SE3STCH_VL3}
\end{align}
Due to the fact that ${\rm Tr}\{\hat{R}\left[\xi\right]_{{\rm D}}\hat{R}^{\top}\}=\sum_{i}^{3}\xi$,
define $\bar{\boldsymbol{\xi}}=\sum_{i}^{3}\xi_{i}$. As such, one
may obtain
\[
\frac{1}{2}{\rm Tr}\left\{ \left(\left\Vert \mathcal{E}_{P}\right\Vert ^{2}\mathbf{I}_{3}+2\mathcal{E}_{P}\mathcal{E}_{P}^{\top}\right)\left[\xi\right]_{{\rm D}}\right\} \leq\frac{3}{2}\left\Vert \mathcal{E}_{P}\right\Vert ^{2}\bar{\boldsymbol{\xi}}
\]
Combining the above expression with the Young’s inequality produces
the following results
\begin{align}
\frac{3}{2}\left\Vert \mathcal{E}_{P}\right\Vert ^{2}\bar{\xi} & \leq\frac{9}{8\varrho}\left\Vert \mathcal{E}_{P}\right\Vert ^{4}+\frac{\varrho}{2}\bar{\boldsymbol{\xi}}^{2}\label{eq:SE3STCH_Ry_Young2}
\end{align}
with $\varrho$ being a small positive constant. Combining \eqref{eq:SE3STCH_Ry_Young2}
with \eqref{eq:SE3STCH_VL3} yields
\begin{align}
\mathcal{L}V\leq & \frac{1}{2}(1+\mathcal{E}_{R})\exp(\mathcal{E}_{R})\boldsymbol{\Upsilon}_{a}^{\top}(\tilde{R})\hat{R}(\tilde{b}_{\Omega}-W_{\Omega})\nonumber \\
& +\frac{3}{8}(2+\mathcal{E}_{R})\exp(\mathcal{E}_{R})\boldsymbol{\Upsilon}_{a}^{\top}(\tilde{R})\hat{R}\left[\sigma\right]_{{\rm D}}\hat{R}^{\top}\boldsymbol{\Upsilon}_{a}(\tilde{R})\nonumber \\
& +\left\Vert \mathcal{E}_{P}\right\Vert ^{2}\mathcal{E}_{P}^{\top}\left([\hat{P}]_{\times}\hat{R}(\tilde{b}_{\Omega}-W_{\Omega})+\hat{R}(\tilde{b}_{V}-W_{V})\right)\nonumber \\
& +\frac{9}{8\varrho}\left\Vert \mathcal{E}_{P}\right\Vert ^{4}+\frac{\varrho}{2}\bar{\boldsymbol{\xi}}^{2}-\frac{1}{\gamma_{b}}\tilde{b}^{\top}\dot{\hat{b}}-\frac{1}{\gamma_{\sigma}}\tilde{\sigma}^{\top}\dot{\hat{\sigma}}\label{eq:SE3STCH_VL4}
\end{align}
With direct substitution for the correction factor $W_{\Omega}$ and
$W_{V}$ in \eqref{eq:SE3STCH_Wy_om} and \eqref{eq:SE3STCH_Wy_v},
respectively, and the differential operators $\dot{\hat{b}}$ and
$\dot{\hat{\sigma}}$ in \eqref{eq:SE3STCH_by_om}, \eqref{eq:SE3STCH_by_v},
and \eqref{eq:SE3STCH_sy_om}, respectively, into \eqref{eq:SE3STCH_VL4}
yields
\begin{align}
\mathcal{L}V\leq & -\left(k_{w}-\frac{3}{4}\right)\frac{1+\mathcal{E}_{R}}{1-\mathcal{E}_{R}}\exp\left(\mathcal{E}_{R}\right)\nonumber \\
& \hspace{8em}\times\boldsymbol{\Upsilon}_{a}^{\top}(\tilde{R})\hat{R}\left[\sigma\right]_{{\rm D}}\hat{R}^{\top}\boldsymbol{\Upsilon}_{a}(\tilde{R})\nonumber \\
& -\frac{1}{\varrho}\left(k_{w}-\frac{9}{8}\right)\left\Vert \mathcal{E}_{P}\right\Vert ^{4}-k_{b}\left\Vert \tilde{b}\right\Vert ^{2}-k_{\sigma}\left\Vert \tilde{\sigma}\right\Vert ^{2}\nonumber \\
& +k_{b}\tilde{b}^{\top}b+k_{\sigma}\tilde{\sigma}^{\top}\sigma+\frac{\varrho}{2}\bar{\xi}^{2}\label{eq:SE3STCH_VL5}
\end{align}
which implies that
\begin{align}
\mathcal{L}V\leq & -\frac{4k_{w}-3}{4}\underline{\lambda}\left(\left[\sigma\right]_{{\rm D}}\right)\frac{1+\mathcal{E}_{R}}{1-\mathcal{E}_{R}}\exp\left(\mathcal{E}_{R}\right)\left\Vert \boldsymbol{\Upsilon}_{a}(\tilde{R})\right\Vert ^{2}\nonumber \\
& -\frac{1}{\varrho}\frac{8k_{w}-9}{8}\left\Vert \mathcal{E}_{P}\right\Vert ^{4}-k_{b}\left\Vert \tilde{b}\right\Vert ^{2}-k_{\sigma}\left\Vert \tilde{\sigma}\right\Vert ^{2}\nonumber \\
& +k_{b}\tilde{b}^{\top}b+k_{\sigma}\tilde{\sigma}^{\top}\sigma+\frac{\varrho}{2}\bar{\xi}^{2}\label{eq:SE3STCH_VL6}
\end{align}
From \eqref{eq:SE3PPF_lemm1_1} in Lemma \ref{Lemm:SE3STCH_1} it
follows that $||\boldsymbol{\Upsilon}_{a}(\tilde{R})||^{2}=4(1-\mathcal{E}_{R})\mathcal{E}_{R}$
and $\underline{\lambda}(\left[\sigma\right]_{{\rm D}})$ refers to
the minimum value of $\sigma$. In view of Young’s inequality, one
obtains
\begin{align*}
k_{b}\tilde{b}^{\top}b & \leq\frac{k_{b}}{2}\left\Vert b\right\Vert ^{2}+\frac{k_{b}}{2}\left\Vert \tilde{b}\right\Vert ^{2}\\
k_{\sigma}\tilde{\sigma}^{\top}\sigma & \leq\frac{k_{\sigma}}{2}\left\Vert \sigma\right\Vert ^{2}+\frac{k_{\sigma}}{2}\left\Vert \tilde{\sigma}\right\Vert ^{2}
\end{align*}
Consequently, the result in \eqref{eq:SE3STCH_VL6} becomes
\begin{align}
\mathcal{L}V\leq & -\left(4k_{w}-3\right)\underline{\lambda}(\left[\sigma\right]_{{\rm D}})(1+\mathcal{E}_{R})\exp(\mathcal{E}_{R})\mathcal{E}_{R}\nonumber \\
& -\frac{1}{\varrho}\frac{8k_{w}-9}{8}\left\Vert \mathcal{E}_{P}\right\Vert ^{4}-\frac{k_{b}}{2}\left\Vert \tilde{b}\right\Vert ^{2}-\frac{k_{\sigma}}{2}\left\Vert \tilde{\sigma}\right\Vert ^{2}\nonumber \\
& +\frac{k_{b}}{2}\left\Vert b\right\Vert ^{2}+\frac{k_{\sigma}}{2}\left\Vert \sigma\right\Vert ^{2}+\frac{\varrho}{2}\bar{\boldsymbol{\xi}}^{2}\label{eq:SE3STCH_VL7}
\end{align}
Recall that $b$ and $\sigma$ are bounded as defined in Assumption
\ref{Assum:SE3STCH_2}. Setting $\gamma_{b},\gamma_{\sigma},k_{b},k_{\sigma}>0$,
$k_{w}>9/8$, and the positive constant $\varrho$ sufficiently small,
the operator $\mathcal{L}V$ in \eqref{eq:SE3STCH_VL7} becomes similar
to \eqref{eq:SE3PPF_lemm1_1} in Lemma \ref{lem:SE3STCH_1}. Define
$\underline{\lambda}_{2}=\underline{\lambda}(\left[\sigma\right]_{{\rm D}})$,
\[
\mathbf{k}=\frac{k_{b}}{2}\left\Vert b\right\Vert ^{2}+\frac{k_{\sigma}}{2}\left\Vert \sigma\right\Vert ^{2}+\frac{\varrho}{2}\bar{\boldsymbol{\xi}}^{2}
\]
and
\[
\mathcal{H}=\left[\begin{array}{cccc}
(4k_{w}-3)\underline{\lambda}_{2} & \mathbf{0}_{1\times3} & \mathbf{0}_{1\times6} & \mathbf{0}_{1\times3}\\
\mathbf{0}_{3\times1} & \frac{1}{\varrho}\frac{8k_{w}-9}{2}\mathbf{I}_{3} & \mathbf{0}_{3\times6} & \mathbf{0}_{3\times3}\\
\mathbf{0}_{6\times1} & \mathbf{0}_{6\times3} & \gamma_{b}k_{b}\mathbf{I}_{6} & \mathbf{0}_{6\times3}\\
\mathbf{0}_{3\times1} & \mathbf{0}_{3\times3} & \mathbf{0}_{3\times6} & \gamma_{\sigma}k_{\sigma}\mathbf{I}_{3}
\end{array}\right]
\]
where $\mathcal{H}\in\mathbb{R}^{13\times13}$. Thereby, the differential
operator in \eqref{eq:SE3STCH_VL7} is equivalent to
\begin{align}
\mathcal{L}V\leq & -\underline{\lambda}\left(\mathcal{H}\right)V+\mathbf{k}\label{eq:SE3STCH_VL_Final}
\end{align}
with $\underline{\lambda}\left(\mathcal{H}\right)$ being the minimum
eigenvalue of $\mathcal{H}$. As such, it can be found that
\begin{align*}
\frac{d\left(\mathbb{E}\left[V\right]\right)}{dt}=\mathbb{E}\left[\mathcal{L}V\right]\leq & -\underline{\lambda}\left(\mathcal{H}\right)\mathbb{E}\left[V\right]+\mathbf{k}
\end{align*}
such that $\frac{d\left(\mathbb{E}\left[V\right]\right)}{dt}\leq0\forall\mathbb{E}\left[V\right]\geq\frac{\mathbf{k}}{\underline{\lambda}\left(\mathcal{H}\right)}$.
Thus, in consistence with Lemma \ref{lem:SE3STCH_1}, the following
result is obtained
\begin{align}
0\leq\mathbb{E}\left[V\right]\leq & V\left(0\right)\exp\left(-\underline{\lambda}\left(\mathcal{H}\right)t\right)+\frac{\mathbf{k}}{\underline{\lambda}\left(\mathcal{H}\right)},\forall t\geq0\label{eq:SE3STCH_V_Final}
\end{align}
Considering that $\boldsymbol{{\rm Y}}=[\mathcal{E}^{\top},\tilde{b}^{\top},\tilde{\sigma}^{\top}]^{\top}\in\mathbb{R}^{13}$
and bearing in mind the result in \eqref{eq:SE3STCH_V_Final}, it
can be easily shown that $\mathbb{E}\left[V\right]$ is eventually
ultimately bounded by $\mathbf{k}/\underline{\lambda}\left(\mathcal{H}\right)$.
Accordingly, $\boldsymbol{{\rm Y}}$ is SGUUB in the mean square.
For a rotation matrix $\tilde{R}\in\mathbb{SO}\left(3\right)$, $\mathcal{E}_{R}$
and $\mathcal{E}_{P}$, define $\mathcal{U}_{0}\subseteq\mathbb{R}\times\mathbb{R}^{3}\times\mathbb{R}^{6}\times\mathbb{R}^{3}$
such that $\mathcal{U}_{0}=\left\{ (\tilde{R}_{0},\tilde{P}_{0},\tilde{b}_{0},\tilde{\sigma}_{0})|\mathcal{E}_{R}\left(0\right)=+1,\tilde{P}_{0}=0,\tilde{b}_{0}=0,\tilde{\sigma}_{0}=0\right\} $.
The set $\mathcal{U}_{0}$ is forward invariant and unstable for the
pose dynamics in \eqref{eq:SE3PPF_T_Dynamics}. Thus, from almost
any initial condition that satisfies $\mathcal{E}_{R}\left(0\right)\notin\mathcal{U}_{0}$,
or equivalently, ${\rm Tr}\{\tilde{R}_{0}\}\neq-1$, the trajectory
of $\boldsymbol{{\rm Y}}$ is SGUUB in the mean square.

\subsection{Direct Nonlinear Stochastic Pose Estimator on $\mathbb{SE}\left(3\right)$
	\label{subsec:SE3_Direct-Filter}}

The reconstructed matrix $\boldsymbol{T}_{y}$ given in Subsection
\ref{subsec:SE3_Passive-Filter} contains two elements: $R_{y}$ and
$P_{y}$. In spite of the fact that $R_{y}$ can be easily reconstructed,
for instance, through QUEST \cite{shuster1981three}, or SVD \cite{markley1988attitude},
the previously proposed methods of static reconstruction increase
the processing cost \cite{mohamed2019filters,hashim2018SO3Stochastic}.
The nonlinear stochastic estimator introduced in this Subsection circumvents
the need for $R_{y}$ reconstruction by directly utilizing the measurements
obtained from the inertial and body-frame units. Consider
\begin{align}
\boldsymbol{\mathcal{M}}_{{\rm T}}=\left[\begin{array}{cc}
\mathbf{M}_{{\rm T}} & \mathbf{m}_{{\rm v}}\\
\mathbf{m}_{{\rm v}}^{\top} & \mathbf{m}_{{\rm c}}
\end{array}\right]= & \sum_{i=1}^{N_{{\rm R}}}s_{i}^{{\rm R}}\left[\begin{array}{c}
\upsilon_{i}^{\mathcal{I}\left({\rm R}\right)}\\
0
\end{array}\right]\left[\begin{array}{c}
\upsilon_{i}^{\mathcal{I}\left({\rm R}\right)}\\
0
\end{array}\right]^{\top}+\nonumber \\
& \sum_{j=1}^{N_{{\rm L}}}s_{j}^{{\rm L}}\left[\begin{array}{c}
{\rm v}_{j}^{\mathcal{I}\left({\rm L}\right)}\\
1
\end{array}\right]\left[\begin{array}{c}
{\rm v}_{j}^{\mathcal{I}\left({\rm L}\right)}\\
1
\end{array}\right]^{\top}\label{eq:SE3PPF_MI}
\end{align}
with $s_{i}^{{\rm R}}\geq0$ and $s_{j}^{{\rm L}}\geq0$ being the
constant gains associated with the confidence level of the $i$th
and $j$th sensor measurements, respectively, also, $\mathbf{M}_{{\rm T}}=\mathbf{M}_{{\rm R}}+\mathbf{M}_{{\rm L}}$
and
\begin{equation}
\begin{cases}
\mathbf{M}_{{\rm R}}= & \sum_{i=1}^{N_{{\rm R}}}s_{i}^{{\rm R}}\upsilon_{i}^{\mathcal{I}\left({\rm R}\right)}\left(\upsilon_{i}^{\mathcal{I}\left({\rm R}\right)}\right)^{\top}\\
\mathbf{M}_{{\rm L}}= & \sum_{j=1}^{N_{{\rm L}}}s_{j}^{{\rm L}}{\rm v}_{j}^{\mathcal{I}\left({\rm L}\right)}\left({\rm v}_{j}^{\mathcal{I}\left({\rm L}\right)}\right)^{\top}\\
\mathbf{m}_{{\rm v}}= & \sum_{j=1}^{N_{{\rm L}}}s_{j}^{{\rm L}}{\rm v}_{j}^{\mathcal{I}\left({\rm L}\right)}\\
\mathbf{m}_{{\rm c}}= & \sum_{j=1}^{N_{{\rm L}}}s_{j}^{{\rm L}}
\end{cases}\label{eq:SE3PPF_MI_elements}
\end{equation}
Also, define
\begin{align}
\boldsymbol{\mathcal{K}}_{{\rm T}}=\left[\begin{array}{cc}
\mathbf{K}_{{\rm T}} & \mathbf{k}_{{\rm v}}\\
\mathbf{m}_{{\rm v}}^{\top} & \mathbf{m}_{{\rm c}}
\end{array}\right]= & \sum_{i=1}^{N_{{\rm R}}}s_{i}^{{\rm R}}\left[\begin{array}{c}
\upsilon_{i}^{\mathcal{B}\left({\rm R}\right)}\\
0
\end{array}\right]\left[\begin{array}{c}
\upsilon_{i}^{\mathcal{I}\left({\rm R}\right)}\\
0
\end{array}\right]^{\top}+\nonumber \\
& \sum_{j=1}^{N_{{\rm L}}}s_{j}^{{\rm L}}\left[\begin{array}{c}
{\rm v}_{j}^{\mathcal{B}\left({\rm L}\right)}\\
1
\end{array}\right]\left[\begin{array}{c}
{\rm v}_{j}^{\mathcal{I}\left({\rm L}\right)}\\
1
\end{array}\right]^{\top}\label{eq:SE3PPF_KIB}
\end{align}
where $\mathbf{m}_{{\rm v}}=\sum_{j=1}^{N_{{\rm L}}}s_{j}^{{\rm L}}{\rm v}_{j}^{\mathcal{I}\left({\rm L}\right)}$,
$\mathbf{m}_{{\rm c}}=\sum_{j=1}^{N_{{\rm L}}}s_{j}^{{\rm L}}$, and
\begin{equation}
\begin{cases}
\mathbf{K}_{{\rm T}}= & \sum_{i=1}^{N_{{\rm R}}}s_{i}^{{\rm R}}\upsilon_{i}^{\mathcal{B}\left({\rm R}\right)}\left(\upsilon_{i}^{\mathcal{I}\left({\rm R}\right)}\right)^{\top}\\
& +\sum_{j=1}^{N_{{\rm L}}}s_{i}^{{\rm L}}{\rm v}_{j}^{\mathcal{B}\left({\rm L}\right)}\left({\rm v}_{j}^{\mathcal{I}\left({\rm L}\right)}\right)^{\top}\\
\mathbf{k}_{{\rm v}}= & \sum_{j=1}^{N_{{\rm L}}}s_{j}^{{\rm L}}{\rm v}_{j}^{\mathcal{B}\left({\rm L}\right)}
\end{cases}\label{eq:SE3PPF_KIB_elements-1}
\end{equation}
It is worth mentioning that $s_{i}^{{\rm R}}$ is selected such that
it satisfies $\sum_{i=1}^{N_{{\rm R}}}s_{i}^{{\rm R}}=3$ and $\sum_{j=1}^{N_{{\rm L}}}s_{j}^{{\rm L}}\neq0$.
It is clear that $\mathbf{M}_{{\rm R}}$ is symmetric. Letting Assumption
\ref{Assum:SE3STCH_1} hold implies that $\mathbf{M}_{{\rm R}}$ is
nonsingular with ${\rm rank}(\mathbf{M}_{{\rm R}})=3$. Define the
three eigenvalues of $\mathbf{M}_{{\rm R}}$ as $\lambda(\mathbf{M}_{{\rm R}})=\{\lambda_{1},\lambda_{2},\lambda_{3}\}$.
Thereby, $\lambda_{1}$, $\lambda_{2}$, and $\lambda_{3}$ are greater
than zero. Let $\bar{\mathbf{M}}_{{\rm R}}={\rm Tr}\{\mathbf{M}_{{\rm R}}\}\mathbf{I}_{3}-\mathbf{M}_{{\rm R}}\in\mathbb{R}^{3\times3}$,
provided that ${\rm rank}(\mathbf{M}_{{\rm R}})=3$. The subsequent
statements are true (\cite{bullo2004geometric} page. 553): 
\begin{enumerate}
	\item The matrix $\mathbf{M}_{{\rm R}}$ is positive-definite.
	\item The eigenvalues of $\mathbf{M}_{{\rm R}}$ are $\lambda(\bar{\mathbf{M}}_{{\rm R}})=\{\lambda_{3}+\lambda_{2},\lambda_{3}+\lambda_{1},\lambda_{2}+\lambda_{1}\}$
	with $\underline{\lambda}(\bar{\mathbf{M}}_{{\rm R}})>0$ being the
	minimum singular value of the set. 
\end{enumerate}
To guarantee that these two statements remain true, it is considered
that ${\rm rank}(\mathbf{M}_{{\rm R}})=3$ in the rest of this subsection.
Let 
\begin{equation}
\hat{\upsilon}_{i}^{\mathcal{B}\left({\rm R}\right)}=\hat{R}^{\top}\upsilon_{i}^{\mathcal{I}\left({\rm R}\right)}\label{eq:SE3PPF_vB_hat}
\end{equation}
Define the homogeneous transformation matrix error $\tilde{\boldsymbol{T}}=\hat{\boldsymbol{T}}\boldsymbol{T}^{-1}$
as in \eqref{eq:SE3STCH_T_error}. It follows that $\tilde{R}=\hat{R}R^{\top}$
and $\tilde{P}=\hat{P}-\tilde{R}P$. Define the error in $b$ and
$\sigma$ as in \eqref{eq:SE3STCH_bt} and \eqref{eq:SE3STCH_st},
respectively. To introduce the direct stochastic pose estimator on
$\mathbb{SE}\left(3\right)$, it is necessary to define a set of expressions
in terms of vector measurements. Therefore, let us define the following
terms: $\mathbf{vex}(\boldsymbol{\mathcal{P}}_{a}(\tilde{R}\mathbf{M}_{{\rm R}}))$,
$\tilde{R}\mathbf{M}_{{\rm R}}$, $||\tilde{R}\mathbf{M}_{{\rm R}}||_{I}$,
and $\tilde{P}$. From the identities in \eqref{eq:SE3STCH_Identity1}
and \eqref{eq:SE3STCH_Identity2}, one obtains
\begin{align*}
& \left[\hat{R}\sum_{i=1}^{N_{{\rm R}}}\frac{s_{i}^{{\rm R}}}{2}\hat{\upsilon}_{i}^{\mathcal{B}\left({\rm R}\right)}\times\upsilon_{i}^{\mathcal{B}\left({\rm R}\right)}\right]_{\times}\\
& =\hat{R}\sum_{i=1}^{N_{{\rm R}}}\frac{s_{i}^{{\rm R}}}{2}\left(\upsilon_{i}^{\mathcal{B}\left({\rm R}\right)}\left(\hat{\upsilon}_{i}^{\mathcal{B}\left({\rm R}\right)}\right)^{\top}-\hat{\upsilon}_{i}^{\mathcal{B}\left({\rm R}\right)}\left(\upsilon_{i}^{\mathcal{B}\left({\rm R}\right)}\right)^{\top}\right)\hat{R}^{\top}\\
& =\frac{1}{2}\hat{R}R^{\top}\mathbf{M}_{{\rm R}}-\frac{1}{2}\mathbf{M}_{{\rm R}}R\hat{R}^{\top}\\
& =\boldsymbol{\mathcal{P}}_{a}(\tilde{R}\mathbf{M}_{{\rm R}})
\end{align*}
which implies that
\begin{equation}
\boldsymbol{\Upsilon}_{a}(\tilde{R}\mathbf{M}_{{\rm R}})=\hat{R}\sum_{i=1}^{N_{{\rm R}}}\left(\frac{s_{i}^{{\rm R}}}{2}\hat{\upsilon}_{i}^{\mathcal{B}\left({\rm R}\right)}\times\upsilon_{i}^{\mathcal{B}\left({\rm R}\right)}\right)\label{eq:SE3PPF_VEX_VM}
\end{equation}
where $\boldsymbol{\Upsilon}_{a}(\tilde{R}\mathbf{M}_{{\rm R}})=\mathbf{vex}(\boldsymbol{\mathcal{P}}_{a}(\tilde{R}\mathbf{M}_{{\rm R}}))$.
Therefore, $\tilde{R}\mathbf{M}_{{\rm R}}$ could be expressed with
respect to vector measurements as
\begin{equation}
\tilde{R}\mathbf{M}_{{\rm R}}=\hat{R}\sum_{i=1}^{N_{{\rm R}}}\left(s_{i}^{{\rm R}}\upsilon_{i}^{\mathcal{B}\left({\rm R}\right)}\left(\upsilon_{i}^{\mathcal{I}\left({\rm R}\right)}\right)^{\top}\right)\label{eq:SE3PPF_RM_VM}
\end{equation}
Hence, the normalized Euclidean distance of \eqref{eq:SE3PPF_RM_VM}
is
\begin{align}
||\tilde{R}\mathbf{M}_{{\rm R}}||_{{\rm I}} & =\frac{1}{4}{\rm Tr}\left\{ (\mathbf{I}_{3}-\tilde{R})\mathbf{M}_{{\rm R}}\right\} \nonumber \\
& =\frac{1}{4}{\rm Tr}\left\{ \mathbf{I}_{3}-\hat{R}\sum_{i=1}^{N_{{\rm R}}}\left(k_{i}^{{\rm R}}\upsilon_{i}^{\mathcal{B}\left({\rm R}\right)}\left(\upsilon_{i}^{\mathcal{I}\left({\rm R}\right)}\right)^{\top}\right)\right\} \nonumber \\
& =\frac{1}{4}\sum_{i=1}^{N_{{\rm R}}}\left(1-\left(\hat{\upsilon}_{i}^{\mathcal{B}\left({\rm R}\right)}\right)^{\top}\upsilon_{i}^{\mathcal{B}\left({\rm R}\right)}\right)\label{eq:SE3PPF_RI_VM}
\end{align}
From \nameref{sec:SO3_PPF_STCH_AppendixA}, it becomes apparent that
\begin{align}
1-||\tilde{R}||_{{\rm I}} & =\frac{1}{4}(1+{\rm Tr}\{\tilde{R}\mathbf{M}_{{\rm R}}\mathbf{M}_{{\rm R}}^{-1}\})\label{eq:SE3PPF_EXPL_append_rho2}
\end{align}
From \eqref{eq:SE3PPF_EXPL_append_rho2}, one has{\small{}
	\begin{align}
	& {\rm Tr}\left\{ \tilde{R}\mathbf{M}_{{\rm R}}\mathbf{M}_{{\rm R}}^{-1}\right\} \nonumber \\
	& ={\rm Tr}\left\{ \left(\sum_{i=1}^{N_{{\rm R}}}s_{i}^{{\rm R}}\upsilon_{i}^{\mathcal{B}\left({\rm R}\right)}\left(\upsilon_{i}^{\mathcal{I}\left({\rm R}\right)}\right)^{\top}\right)\left(\sum_{i=1}^{N_{{\rm R}}}s_{i}^{{\rm R}}\hat{\upsilon}_{i}^{\mathcal{B}\left({\rm R}\right)}\left(\upsilon_{i}^{\mathcal{I}\left({\rm R}\right)}\right)^{\top}\right)^{-1}\right\} \label{eq:SE3PPF_Gamma_VM}
	\end{align}
}From \eqref{eq:SE3PPF_MI} and \eqref{eq:SE3PPF_MI_elements}, one
has
\begin{align}
\tilde{\boldsymbol{T}}\boldsymbol{\mathcal{M}}^{\mathcal{I}} & =\left[\begin{array}{cc}
\tilde{R}\mathbf{M}_{{\rm T}}+\tilde{P}\mathbf{m}_{{\rm v}}^{\top} & \tilde{R}\mathbf{m}_{{\rm v}}+\mathbf{m}_{{\rm c}}\tilde{P}\\
\mathbf{m}_{{\rm v}}^{\top} & \mathbf{m}_{{\rm c}}
\end{array}\right]\label{eq:SE3PPF_VM_Part1}
\end{align}
The expression in \eqref{eq:SE3PPF_VM_Part1} can be transformed as
follows
\begin{align}
\tilde{\boldsymbol{T}}\boldsymbol{\mathcal{M}}^{\mathcal{I}} & =\left[\begin{array}{cc}
\hat{R} & \hat{P}\\
\mathbf{0}_{3\times1}^{\top} & 1
\end{array}\right]\left[\begin{array}{cc}
\mathbf{K}_{{\rm T}} & \mathbf{k}_{{\rm v}}\\
\mathbf{m}_{{\rm v}}^{\top} & \mathbf{m}_{{\rm c}}
\end{array}\right]\nonumber \\
& =\left[\begin{array}{cc}
\hat{R}\mathbf{K}_{{\rm T}}+\hat{P}\mathbf{m}_{{\rm v}}^{\top} & \hat{R}\mathbf{k}_{{\rm v}}+\mathbf{m}_{{\rm c}}\hat{P}\\
\mathbf{m}_{{\rm v}}^{\top} & \mathbf{m}_{{\rm c}}
\end{array}\right]\label{eq:SE3PPF_VM_Part2}
\end{align}
From \eqref{eq:SE3PPF_VM_Part1} and \eqref{eq:SE3PPF_VM_Part2},
the position error can be evaluated in view of the vector measurements
as
\begin{equation}
\tilde{P}=\hat{P}+\frac{1}{\mathbf{m}_{{\rm c}}}\left(\hat{R}\mathbf{k}_{{\rm v}}-\tilde{R}\mathbf{M}_{{\rm R}}\mathbf{M}_{{\rm R}}^{-1}\mathbf{m}_{{\rm v}}\right)\label{eq:SE3PPF_Ptil_VM}
\end{equation}
where $\tilde{R}\mathbf{M}_{{\rm R}}$ is calculated as in \eqref{eq:SE3PPF_RM_VM}.
As such, in all the subsequent derivations and calculations $\mathbf{vex}(\boldsymbol{\mathcal{P}}_{a}(\tilde{R}\mathbf{M}_{{\rm R}}))$,
$\tilde{R}\mathbf{M}_{{\rm R}}$, $||\tilde{R}\mathbf{M}_{{\rm R}}||_{{\rm I}}$,
${\rm Tr}\{\tilde{R}\mathbf{M}_{{\rm R}}\mathbf{M}_{{\rm R}}^{-1}\}$,
and $\tilde{P}$ are extracted via a set of vector measurements as
defined in \eqref{eq:SE3PPF_VEX_VM}, \eqref{eq:SE3PPF_RM_VM}, \eqref{eq:SE3PPF_RI_VM},
\eqref{eq:SE3PPF_Gamma_VM}, and \eqref{eq:SE3PPF_Ptil_VM}, respectively.
Modify the vector error in \eqref{eq:SE3STCH_E1} and redefine it
as follows
\begin{equation}
\mathcal{E}=\left[\mathcal{E}_{R},\mathcal{E}_{P}^{\top}\right]^{\top}=\left[||\tilde{R}\mathbf{M}_{{\rm R}}||_{{\rm I}},\tilde{P}^{\top}\right]^{\top}\label{eq:SE3PPF_E_VM}
\end{equation}
where $\mathcal{E}_{R}=||\tilde{R}\mathbf{M}_{{\rm R}}||_{{\rm I}}$
and $\mathcal{E}_{P}=\tilde{P}$ are defined in \eqref{eq:SE3PPF_RI_VM}
and \eqref{eq:SE3PPF_Ptil_VM}, respectively. Consider the following
estimator design
\begin{equation}
\left[\begin{array}{cc}
\dot{\hat{R}} & \dot{\hat{P}}\\
\mathbf{0}_{3\times1}^{\top} & 0
\end{array}\right]=\left[\begin{array}{cc}
\hat{R} & \hat{P}\\
\mathbf{0}_{3\times1}^{\top} & 1
\end{array}\right]\left[\begin{array}{c}
\Omega_{m}-\hat{b}_{\Omega}-W_{\Omega}\\
V_{m}-\hat{b}_{V}-W_{V}
\end{array}\right]_{\land}\label{eq:SE3PPF_Tvm_dot}
\end{equation}
\begin{align}
W_{\Omega}= & \frac{4}{\underline{\lambda}_{1}}\frac{k_{w}\left[\hat{R}^{\top}\boldsymbol{\Upsilon}_{a}(\tilde{R}\mathbf{M}_{{\rm R}})\right]_{{\rm D}}}{1+{\rm Tr}\{\tilde{R}\mathbf{M}_{{\rm R}}\mathbf{M}_{{\rm R}}^{-1}\}}\hat{\sigma}\label{eq:SE3STCH_Wvm_om}\\
W_{V}= & -\hat{R}^{\top}\left[\hat{P}\right]_{\times}\hat{R}W_{\Omega}+\frac{k_{w}}{\varrho}\hat{R}^{\top}\mathcal{E}_{P}\label{eq:SE3STCH_Wvm_v}\\
\dot{\hat{b}}_{\Omega}= & \frac{\gamma_{b}}{2}\left(1+\mathcal{E}_{R}\right)\exp\left(\mathcal{E}_{R}\right)\hat{R}^{\top}\boldsymbol{\Upsilon}_{a}(\tilde{R}\mathbf{M}_{{\rm R}})\nonumber \\
& -\gamma_{b}\left\Vert \mathcal{E}_{P}\right\Vert ^{2}\hat{R}^{\top}\left[\hat{P}\right]_{\times}\mathcal{E}_{P}-\gamma_{b}k_{b}\hat{b}_{\Omega}\label{eq:SE3STCH_bvm_om}\\
\dot{\hat{b}}_{V}= & \gamma_{b}\left\Vert \mathcal{E}_{P}\right\Vert ^{2}\hat{R}^{\top}\mathcal{E}_{P}-\gamma_{b}k_{b}\hat{b}_{V}\label{eq:SE3STCH_bvm_v}\\
K_{\mathcal{E}}= & \gamma_{\sigma}\frac{1+\mathcal{E}_{R}}{1+{\rm Tr}\{\tilde{R}\mathbf{M}_{{\rm R}}\mathbf{M}_{{\rm R}}^{-1}\}}\exp\left(\mathcal{E}_{R}\right)\nonumber \\
\dot{\hat{\sigma}}= & \frac{2k_{w}}{\underline{\lambda}_{1}}K_{\mathcal{E}}\left[\hat{R}^{\top}\boldsymbol{\Upsilon}_{a}(\tilde{R}\mathbf{M}_{{\rm R}})\right]_{{\rm D}}\hat{R}^{\top}\boldsymbol{\Upsilon}_{a}(\tilde{R}\mathbf{M}_{{\rm R}})\nonumber \\
& -\gamma_{\sigma}k_{\sigma}\hat{\sigma}\label{eq:SE3STCH_svm_om}
\end{align}
with $\left[\mathcal{E}_{R},\mathcal{E}_{P}^{\top}\right]^{\top}=\left[||\tilde{R}\mathbf{M}_{{\rm R}}||_{{\rm I}},\tilde{P}^{\top}\right]^{\top}$
and $\boldsymbol{\Upsilon}_{a}(\tilde{R}\mathbf{M}_{{\rm R}})$ being
specified in \eqref{eq:SE3PPF_RI_VM}, \eqref{eq:SE3PPF_Ptil_VM}
and \eqref{eq:SE3PPF_VEX_VM}, respectively. $\left[\cdot\right]_{{\rm D}}$
is a diagonal matrix of the associated vector, $\underline{\lambda}_{1}=\underline{\lambda}(\bar{\mathbf{M}}_{{\rm R}})$
is the minimum singular value of $\bar{\mathbf{M}}_{{\rm R}}$, $k_{w}$,
$\gamma_{b}$, and $\gamma_{\sigma}$ are positive constants, while
$\hat{b}=\left[\hat{b}_{\Omega}^{\top},\hat{b}_{V}^{\top}\right]^{\top}$
and $\hat{\sigma}$ are the estimates of $b$ and $\sigma$, respectively. The equivalent quaternion representation and complete implementation
steps of the direct filter are given in \nameref{sec:SO3_PPF_STCH_AppendixB}.
\begin{thm}
	\textbf{\label{thm:SE3PPF_2}} Consider the pose estimator in \eqref{eq:SE3PPF_Tvm_dot},
	\eqref{eq:SE3STCH_Wvm_om}, \eqref{eq:SE3STCH_Wvm_v}, \eqref{eq:SE3STCH_bvm_om},
	\eqref{eq:SE3STCH_bvm_v}, and \eqref{eq:SE3STCH_svm_om} geared with
	the vector measurements in \eqref{eq:SE3STCH_Set_R_Norm} and \eqref{eq:SE3STCH_Set_L},
	and the velocity measurements in \eqref{eq:SE3PPF_Angular} and \eqref{eq:SE3PPF_V_Trans}.
	Let Assumption \ref{Assum:SE3STCH_1} hold and assume that the selected
	parameters fulfill the following conditions: $\gamma_{b}>0$, $\gamma_{\sigma}>0$,
	$k_{b}>0$, $k_{\sigma}>0$, and $k_{w}>9/8$. Let $\varrho>0$ be
	selected sufficiently small. Consider the set in Remark \ref{rem:Unstable-set}.
	In the event of $\mathcal{Y}_{m}$ are corrupted with unknown constant
	bias and random noise $\left(\omega\neq0\right)$, and $\tilde{R}\left(0\right)\notin\mathcal{S}_{0}$,
	the vector $\left[\mathcal{E}^{\top},\tilde{b}^{\top},\tilde{\sigma}^{\top}\right]^{\top}$
	is semi-globally uniformly ultimately bounded in mean square. Additionally,
	the filter errors could be minimized by the appropriated selection
	of the design parameters.
\end{thm}
\textbf{Proof. }Let the error of $\boldsymbol{T}$, $b$, and $\sigma$
be defined as in \eqref{eq:SE3STCH_T_error}, \eqref{eq:SE3STCH_bt},
and \eqref{eq:SE3STCH_st}, respectively. As such, the error in attitude
dynamics is analogous to \eqref{eq:SE3STCH_Rt_dot}. $\dot{\mathbf{M}}_{{\rm R}}=\mathbf{0}_{3\times3}$
due to the $i$th inertial vector ${\rm v}_{i}^{\mathcal{I}\left({\rm R}\right)}$
being constant. Hence, from \eqref{eq:SE3STCH_Rt_dot}, the derivative
of $||\tilde{R}\mathbf{M}_{{\rm R}}||_{I}$ becomes
\begin{align}
& \frac{d}{dt}||\tilde{R}\mathbf{M}_{{\rm R}}||_{{\rm I}}\nonumber \\
& =-\frac{1}{4}{\rm Tr}\left\{ \left[(\hat{R}\tilde{b}_{\Omega}-W_{\Omega})dt+\hat{R}\mathcal{Q}_{\Omega}d\beta_{\Omega}\right]_{\times}\tilde{R}\mathbf{M}_{{\rm R}}\right\} \nonumber \\
& =\frac{1}{2}\boldsymbol{\Upsilon}_{a}^{\top}(\tilde{R}\mathbf{M}_{{\rm R}})\left((\hat{R}\tilde{b}_{\Omega}-W_{\Omega})dt+\hat{R}\mathcal{Q}_{\Omega}d\beta_{\Omega}\right)\label{eq:SE3PPF_NormMIRtilde_dot}
\end{align}
with ${\rm Tr}\{\left[W_{\Omega}\right]_{\times}\tilde{R}\mathbf{M}_{{\rm R}}\}=-2\mathbf{vex}\left(\boldsymbol{\mathcal{P}}_{a}(\tilde{R}\mathbf{M}_{{\rm R}})\right)^{\top}W_{\Omega}$
as defined in identity \eqref{eq:SE3STCH_Identity6}. It can be demonstrated
that the derivative of $\tilde{P}$ in incremental form is identical
to \eqref{eq:SE3STCH_Pt_dot}. As such, one has
\begin{align}
d\mathcal{E}= & \tilde{\mathcal{F}}dt+\tilde{\mathcal{G}}\mathcal{Q}d\beta\nonumber \\
\tilde{\mathcal{G}}= & \left[\begin{array}{cc}
\frac{1}{2}\boldsymbol{\Upsilon}_{a}^{\top}(\tilde{R}\mathbf{M}_{{\rm R}})\hat{R} & \mathbf{0}_{1\times3}\\{}
[\hat{P}-\tilde{P}]_{\times}\hat{R} & \hat{R}
\end{array}\right]\nonumber \\
\tilde{\mathcal{F}}= & \tilde{\mathcal{G}}((\tilde{b}-W)dt+\mathcal{Q}d\beta)\label{eq:SE3PPF_VEC_MI_tilde_dot}
\end{align}
with $\mathcal{E}$ being defined in terms of vector measurements
in \eqref{eq:SE3PPF_E_VM} and $\tilde{b}=[\tilde{b}_{\Omega}^{\top},\tilde{b}_{V}^{\top}]^{\top}$.
Let $V:=V(\mathcal{E},\tilde{b},\tilde{\sigma})$ and consider the
following Lyapunov candidate function
\begin{align}
V= & \mathcal{E}_{R}\exp\left(\mathcal{E}_{R}\right)+\frac{1}{4}\left\Vert \mathcal{E}_{P}\right\Vert ^{4}+\frac{1}{2\gamma_{b}}\left\Vert \tilde{b}\right\Vert ^{2}+\frac{1}{2\gamma_{\sigma}}\left\Vert \tilde{\sigma}\right\Vert ^{2}\label{eq:SE3STCH_V_vm}
\end{align}
It can be proven that the differential operator $\mathcal{L}V$ in
Definition \ref{def:SE3STCH_2} is analogous to \eqref{eq:SE3STCH_VL1}.
Also, $V_{\mathcal{E}}$ and $V_{\mathcal{E}\mathcal{E}}$ are similar
to \eqref{eq:SE3STCH_Vv} and \eqref{eq:SE3STCH_Vvv}, respectively.
Accordingly, $\mathcal{L}V$ is equivalent to
\begin{align}
& \mathcal{L}V=\frac{1}{2}\left(1+\mathcal{E}_{R}\right)\exp\left(\mathcal{E}_{R}\right)\boldsymbol{\Upsilon}_{a}^{\top}(\tilde{R}\mathbf{M}_{{\rm R}})\hat{R}(\tilde{b}_{\Omega}-W_{\Omega})\nonumber \\
& \hspace{1em}+\frac{3}{8}\left(2+\mathcal{E}_{R}\right)\exp\left(\mathcal{E}_{R}\right)\boldsymbol{\Upsilon}_{a}^{\top}(\tilde{R}\mathbf{M}_{{\rm R}})\hat{R}\mathcal{Q}_{\Omega}^{2}\hat{R}^{\top}\boldsymbol{\Upsilon}_{a}(\tilde{R}\mathbf{M}_{{\rm R}})\nonumber \\
& \hspace{1em}+\left\Vert \mathcal{E}_{P}\right\Vert ^{2}\mathcal{E}_{P}^{\top}[\hat{P}-\tilde{P}]_{\times}\hat{R}(\tilde{b}_{\Omega}-W_{\Omega})\nonumber \\
& \hspace{1em}+\left\Vert \mathcal{E}_{P}\right\Vert ^{2}\mathcal{E}_{P}^{\top}\hat{R}(\tilde{b}_{V}-W_{V})\nonumber \\
& \hspace{1em}+\frac{1}{2}{\rm Tr}\left\{ \left(\left\Vert \mathcal{E}_{P}\right\Vert ^{2}\mathbf{I}_{3}+2\mathcal{E}_{P}\mathcal{E}_{P}^{\top}\right)\hat{R}\mathcal{Q}_{V}^{2}\hat{R}^{\top}\right\} \nonumber \\
& \hspace{1em}+\frac{1}{2}{\rm Tr}\left\{ \left(\left\Vert \mathcal{E}_{P}\right\Vert ^{2}\mathbf{I}_{3}+2\mathcal{E}_{P}\mathcal{E}_{P}^{\top}\right)[\hat{P}-\tilde{P}]_{\times}\hat{R}\mathcal{Q}_{V}^{2}\hat{R}^{\top}\right.\nonumber \\
& \hspace{1em}\hspace{5em}\times\left.[\hat{P}-\tilde{P}]_{\times}^{\top}\right\} -\frac{1}{\gamma_{b}}\tilde{b}^{\top}\dot{\hat{b}}-\frac{1}{\gamma_{\sigma}}\tilde{\sigma}^{\top}\dot{\hat{\sigma}}\label{eq:SE3STCH_VL2-1}
\end{align}
$\hat{R}\mathcal{Q}_{V}^{2}\hat{R}^{\top}$ is positive semi-definite
which means that the last trace component in \eqref{eq:SE3STCH_VL2-1}
is negative semi-definite, and therefore can be disregarded. Taking
into consideration $\tilde{P}^{\top}[\tilde{P}]_{\times}=\mathbf{0}_{1\times3}$,
the differential operator in \eqref{eq:SE3STCH_VL2-1} can be transformed
as follows
\begin{align}
& \mathcal{L}V\leq\frac{1}{2}(1+\mathcal{E}_{R})\exp(\mathcal{E}_{R})\boldsymbol{\Upsilon}_{a}^{\top}(\tilde{R}\mathbf{M}_{{\rm R}})\hat{R}(\tilde{b}_{\Omega}-W_{\Omega})\nonumber \\
& \hspace{1em}+\frac{3}{8}(2+\mathcal{E}_{R})\exp\left(\mathcal{E}_{R}\right)\boldsymbol{\Upsilon}_{a}^{\top}(\tilde{R}\mathbf{M}_{{\rm R}})\hat{R}\mathcal{Q}_{\Omega}^{2}\hat{R}^{\top}\boldsymbol{\Upsilon}_{a}(\tilde{R}\mathbf{M}_{{\rm R}})\nonumber \\
& \hspace{1em}+\left\Vert \mathcal{E}_{P}\right\Vert ^{2}\mathcal{E}_{P}^{\top}\left([\hat{P}]_{\times}\hat{R}(\tilde{b}_{\Omega}-W_{\Omega})+\hat{R}(\tilde{b}_{V}-W_{V})\right)\nonumber \\
& \hspace{1em}+\frac{1}{2}{\rm Tr}\left\{ \left(\left\Vert \mathcal{E}_{P}\right\Vert ^{2}\mathbf{I}_{3}+2\mathcal{E}_{P}\mathcal{E}_{P}^{\top}\right)\hat{R}\left[\xi\right]_{{\rm D}}\hat{R}^{\top}\right\} \nonumber \\
& \hspace{1em}-\frac{1}{\gamma_{b}}\tilde{b}^{\top}\dot{\hat{b}}-\frac{1}{\gamma_{\sigma}}\tilde{\sigma}^{\top}\dot{\hat{\sigma}}\label{eq:SE3STCH_VL3-1}
\end{align}
Due to the fact that ${\rm Tr}\{\hat{R}\left[\xi\right]_{{\rm D}}\hat{R}^{\top}\}=\sum_{i}^{3}\xi$,
and define $\bar{\boldsymbol{\xi}}=\sum_{i}^{3}\xi_{i}$ to obtain
the following
\[
\frac{1}{2}{\rm Tr}\left\{ \left(\left\Vert \mathcal{E}_{P}\right\Vert ^{2}\mathbf{I}_{3}+2\mathcal{E}_{P}\mathcal{E}_{P}^{\top}\right)\left[\xi\right]_{{\rm D}}\right\} \leq\frac{3}{2}\left\Vert \mathcal{E}_{P}\right\Vert ^{2}\bar{\boldsymbol{\xi}}
\]
With the aid of the Young’s inequality, one obtains
\begin{align}
\frac{3}{2}\left\Vert \mathcal{E}_{P}\right\Vert ^{2}\bar{\xi} & \leq\frac{9}{8\varrho}\left\Vert \mathcal{E}_{P}\right\Vert ^{4}+\frac{\varrho}{2}\bar{\boldsymbol{\xi}}^{2}\label{eq:SE3STCH_Ry_Young2-1}
\end{align}
Combining the result in \eqref{eq:SE3STCH_Ry_Young2-1} with \eqref{eq:SE3STCH_VL3-1}
and substituting $W_{\Omega}$, $W_{V}$, $\dot{\hat{b}}$ and $\dot{\hat{\sigma}}$
with their definitions in \eqref{eq:SE3STCH_Wy_om} and \eqref{eq:SE3STCH_Wy_v},
\eqref{eq:SE3STCH_by_om}, \eqref{eq:SE3STCH_by_v}, and \eqref{eq:SE3STCH_sy_om},
respectively, yields
\begin{align}
\mathcal{L}V\leq & -\underline{\lambda}_{2}\left(\frac{2k_{w}}{\underline{\lambda}_{1}}-\frac{3}{4}\right)\frac{\left(1+\mathcal{E}_{R}\right)\exp\left(\mathcal{E}_{R}\right)}{1+{\rm Tr}\{\tilde{R}\mathbf{M}_{{\rm R}}\mathbf{M}_{{\rm R}}^{-1}\}}\nonumber \\
& \hspace{6em}\times\boldsymbol{\Upsilon}_{a}^{\top}(\tilde{R}\mathbf{M}_{{\rm R}})\hat{R}\left[\sigma\right]_{{\rm D}}\hat{R}^{\top}\boldsymbol{\Upsilon}_{a}(\tilde{R}\mathbf{M}_{{\rm R}})\nonumber \\
& -\frac{1}{\varrho}\left(k_{w}-\frac{9}{8}\right)\left\Vert \mathcal{E}_{P}\right\Vert ^{4}-k_{b}\left\Vert \tilde{b}\right\Vert ^{2}-k_{\sigma}\left\Vert \tilde{\sigma}\right\Vert ^{2}\nonumber \\
& +k_{b}\tilde{b}^{\top}b+k_{\sigma}\tilde{\sigma}^{\top}\sigma+\frac{\varrho}{2}\bar{\xi}^{2}\label{eq:SE3STCH_VL5-1}
\end{align}
which results in
\begin{align}
\mathcal{L}V\leq & -\underline{\lambda}_{2}\left(\frac{2k_{w}}{\underline{\lambda}_{1}}-\frac{3}{4}\right)\frac{\left(1+\mathcal{E}_{R}\right)\exp\left(\mathcal{E}_{R}\right)\left\Vert \boldsymbol{\Upsilon}_{a}(\tilde{R}\mathbf{M}_{{\rm R}})\right\Vert ^{2}}{1+{\rm Tr}\{\tilde{R}\mathbf{M}_{{\rm R}}\mathbf{M}_{{\rm R}}^{-1}\}}\nonumber \\
& -\frac{1}{\varrho}\frac{8k_{w}-9}{8}\left\Vert \mathcal{E}_{P}\right\Vert ^{4}-k_{b}\left\Vert \tilde{b}\right\Vert ^{2}-k_{\sigma}\left\Vert \tilde{\sigma}\right\Vert ^{2}\nonumber \\
& +k_{b}\tilde{b}^{\top}b+k_{\sigma}\tilde{\sigma}^{\top}\sigma+\frac{\varrho}{2}\bar{\xi}^{2}\label{eq:SE3STCH_VL6-1}
\end{align}
where $||\boldsymbol{\Upsilon}_{a}(\tilde{R}\mathbf{M}_{{\rm R}})||^{2}=||\hat{R}^{\top}\boldsymbol{\Upsilon}_{a}(\tilde{R}\mathbf{M}_{{\rm R}})||^{2}$,
while $\underline{\lambda}_{2}=\underline{\lambda}\left(\left[\sigma\right]_{{\rm D}}\right)$
and $\underline{\lambda}_{1}=\lambda(\bar{\mathbf{M}}_{{\rm R}})$
refer to the minimum value of $\left[\sigma\right]_{{\rm D}}$ and
$\bar{\mathbf{M}}_{{\rm R}}={\rm Tr}\{\mathbf{M}_{{\rm R}}\}\mathbf{I}_{3}-\mathbf{M}_{{\rm R}}$,
respectively. According to Young’s inequality, it can be shown that
\begin{align*}
k_{b}\tilde{b}^{\top}b & \leq\frac{k_{b}}{2}\left\Vert b\right\Vert ^{2}+\frac{k_{b}}{2}\left\Vert \tilde{b}\right\Vert ^{2}\\
k_{\sigma}\tilde{\sigma}^{\top}\sigma & \leq\frac{k_{\sigma}}{2}\left\Vert \sigma\right\Vert ^{2}+\frac{k_{\sigma}}{2}\left\Vert \tilde{\sigma}\right\Vert ^{2}
\end{align*}
Also, from \eqref{eq:SE3PPF_lemm1_2} in Lemma \ref{Lemm:SE3STCH_1},
one has $\frac{2}{\underline{\lambda}_{1}}\frac{||\mathbf{vex}\left(\boldsymbol{\mathcal{P}}_{a}\left(\tilde{R}\mathbf{M}_{{\rm R}}\right)\right)||^{2}}{1+{\rm Tr}\{\tilde{R}\mathbf{M}_{{\rm R}}\mathbf{M}_{{\rm R}}^{-1}\}}\geq\mathcal{E}_{R}$.
Thus, the result in \eqref{eq:SE3STCH_VL6-1} can be expressed as
\begin{align}
\mathcal{L}V\leq & -\underline{\lambda}_{2}\frac{8k_{w}-3\underline{\lambda}_{1}}{8}(1+\mathcal{E}_{R})\exp(\mathcal{E}_{R})\mathcal{E}_{R}\nonumber \\
& -\frac{1}{\varrho}\frac{8k_{w}-9}{8}\left\Vert \mathcal{E}_{P}\right\Vert ^{4}-\frac{k_{b}}{2}\left\Vert \tilde{b}\right\Vert ^{2}-\frac{k_{\sigma}}{2}\left\Vert \tilde{\sigma}\right\Vert ^{2}\nonumber \\
& +\frac{k_{b}}{2}\left\Vert b\right\Vert ^{2}+\frac{k_{\sigma}}{2}\left\Vert \sigma\right\Vert ^{2}+\frac{\varrho}{2}\bar{\boldsymbol{\xi}}^{2}\label{eq:SE3STCH_VL7-1}
\end{align}
It it worth mentioning that $b$ and $\sigma$ are bounded as defined
in Assumption \ref{Assum:SE3STCH_2}. Letting $\gamma_{b},\gamma_{\sigma},k_{b},k_{\sigma}>0$,
$k_{w}>\frac{9}{8}$, $k_{w}>\frac{3}{8}\underline{\lambda}_{1}$
and setting $\varrho$ as a sufficiently small positive constant,
leads to the differential operator $\mathcal{L}V$ in \eqref{eq:SE3STCH_VL7-1}
eventually becoming similar to \eqref{eq:SE3PPF_lemm1_1} in Lemma
\ref{lem:SE3STCH_1}. Let
\[
\mathbf{k}=\frac{k_{b}}{2}\left\Vert b\right\Vert ^{2}+\frac{k_{\sigma}}{2}\left\Vert \sigma\right\Vert ^{2}+\frac{\varrho}{2}\bar{\boldsymbol{\xi}}^{2}
\]
and
\[
\mathcal{H}=\left[\begin{array}{cccc}
\underline{\lambda}_{2}\frac{8k_{w}-3\underline{\lambda}_{1}}{8} & \mathbf{0}_{1\times3} & \mathbf{0}_{1\times6} & \mathbf{0}_{1\times3}\\
\mathbf{0}_{3\times1} & \frac{1}{\varrho}\frac{8k_{w}-9}{2}\mathbf{I}_{3} & \mathbf{0}_{3\times6} & \mathbf{0}_{3\times3}\\
\mathbf{0}_{6\times1} & \mathbf{0}_{6\times3} & \gamma_{b}k_{b}\mathbf{I}_{6} & \mathbf{0}_{6\times3}\\
\mathbf{0}_{3\times1} & \mathbf{0}_{3\times3} & \mathbf{0}_{3\times6} & \gamma_{\sigma}k_{\sigma}\mathbf{I}_{3}
\end{array}\right]
\]
where $\mathcal{H}\in\mathbb{R}^{13\times13}$. Accordingly, $\mathcal{L}V$
in \eqref{eq:SE3STCH_VL7-1} can be written as
\begin{align}
\mathcal{L}V\leq & -\underline{\lambda}\left(\mathcal{H}\right)V+\mathbf{k}\label{eq:SE3STCH_VL_Final-1}
\end{align}
where $\underline{\lambda}\left(\mathcal{H}\right)$ is the minimum
eigenvalue of the matrix $\mathcal{H}$. Based on \eqref{eq:SE3STCH_VL_Final-1},
it can be found that
\begin{align}
\frac{d\left(\mathbb{E}\left[V\right]\right)}{dt}=\mathbb{E}\left[\mathcal{L}V\right]\leq & -\underline{\lambda}\left(\mathcal{H}\right)\mathbb{E}\left[V\right]+\mathbf{k}\label{eq:SE3STCH_VL_Final-2}
\end{align}
and according to Lemma \ref{lem:SE3STCH_1} the following inequality
holds
\begin{align}
0\leq\mathbb{E}\left[V\right]\leq & V\left(0\right)\exp(-\underline{\lambda}\left(\mathcal{H}\right)t)+\frac{\mathbf{k}}{\underline{\lambda}\left(\mathcal{H}\right)},\forall t\geq0\label{eq:SE3STCH_V_Final-1}
\end{align}
which signifies that $\mathbb{E}\left[V\right]$ eventually becomes
ultimately bounded by $\mathbf{k}/\underline{\lambda}\left(\mathcal{H}\right)$.
Let $\boldsymbol{{\rm Y}}=[\mathcal{E}^{\top},\tilde{b}^{\top},\tilde{\sigma}^{\top}]^{\top}\in\mathbb{R}^{13}$.
According to the result in \eqref{eq:SE3STCH_V_Final-1}, $\boldsymbol{{\rm Y}}$
is SGUUB in the mean square. For $\tilde{R}\in\mathbb{SO}\left(3\right)$,
define the following forward invariant and unstable set $\mathcal{U}_{0}\subseteq\mathbb{R}\times\mathbb{R}^{3}\times\mathbb{R}^{6}\times\mathbb{R}^{3}$
for the pose dynamics in \eqref{eq:SE3PPF_T_Dynamics} such that {\small{}
	\[
	\mathcal{U}_{0}=\{(\tilde{R}_{0},\tilde{P}_{0},\tilde{b}_{0},\tilde{\sigma}_{0})|\mathcal{E}_{R}\left(0\right)=+1,\tilde{P}_{0}=0,\tilde{b}_{0}=0,\tilde{\sigma}_{0}=0\}
	\]
}From almost any initial condition such that $\mathcal{E}_{R}\left(0\right)\notin\mathcal{U}_{0}$,
the trajectory of $\boldsymbol{{\rm Y}}$ is SGUUB in the mean square.

\section{Simulation results \label{sec:SE3_Simulations}}

This section presents and compares the performance of the two nonlinear
stochastic estimators on $\mathbb{SE}\left(3\right)$ . Both estimators
are tested against high levels of unknown bias and noise attached
to the measurements of the group velocity vector and the body-frame
vectors and against large initialization error. Let us begin by defining
the homogeneous transformation matrix $\boldsymbol{T}$ as in \eqref{eq:SE3PPF_T_Dynamics}.
Consider the angular velocity $\left({\rm rad/sec}\right)$ to be
given by
\[
\Omega=\left[\begin{array}{ccc}
{\rm sin}\left(\frac{t}{2}\right) & 0.7{\rm sin}\left(\frac{t}{4}+\pi\right) & \frac{1}{2}{\rm sin}\left(0.4t+\frac{\pi}{3}\right)\end{array}\right]^{\top}
\]
and the translational velocity to be
\[
V=\left[\begin{array}{ccc}
{\rm sin}\left(\frac{t}{5}\right) & 0.6{\rm sin}\left(\frac{t+\pi}{2}\right) & {\rm sin}\left(0.4t+\frac{\pi}{4}\right)\end{array}\right]\left({\rm m/sec}\right)
\]
with $R\left(0\right)=\mathbf{I}_{3}$ and $P\left(0\right)=\mathbf{0}_{3\times1}$,
respectively, such that $\boldsymbol{T}=\mathbf{I}_{4}$. Let $\Omega_{m}=\Omega+b_{\Omega}+\omega_{\Omega}$
with the unknown constant bias $b_{\Omega}=0.1\left[1,-1,1\right]^{\top}$
and the unknown random noise vector $\omega_{\Omega}$ having zero
mean and standard deviation (STD) $0.15\left({\rm rad/sec}\right)$.
Consider $V_{m}=V+b_{V}+\omega_{V}$ where the unknown constant bias
$b_{V}=0.1\left[2,5,1\right]^{\top}$ and the random noise vector
$\omega_{V}$ has zero mean and ${\rm STD}=0.15\left({\rm m/sec}\right)$.
Assume there is a landmark available for measurement ${\rm v}_{1}^{\mathcal{I}\left({\rm L}\right)}=\left[\frac{1}{2},\sqrt{2},1\right]^{\top}$
and its body-frame measurement is as \eqref{eq:SE3STCH_Vec_Landmark}.
The associated bias is ${\rm b}_{1}^{\mathcal{B}\left({\rm L}\right)}=0.1\left[0.3,0.2,-0.2\right]^{\top}$
and the noise vector $\omega_{1}^{\mathcal{B}\left({\rm L}\right)}$
has zero mean and ${\rm STD}=0.1\left({\rm m/sec}\right)$. To incorporate
uncertain measurements obtained from an IMU module, let us consider
two non-collinear inertial-frame vectors ${\rm v}_{1}^{\mathcal{I}\left({\rm R}\right)}=\frac{1}{\sqrt{3}}\left[1,-1,1\right]^{\top}$
and ${\rm v}_{2}^{\mathcal{I}\left({\rm R}\right)}=\left[0,0,1\right]^{\top}$
and define body-frame vectors ${\rm v}_{1}^{\mathcal{B}\left({\rm R}\right)}$
and ${\rm v}_{2}^{\mathcal{B}\left({\rm R}\right)}$ according to
\eqref{eq:SE3STCH_Vect_R} for $i=1,2$. The bias associated with
the two body-frame measurements are ${\rm b}_{1}^{\mathcal{B}\left({\rm R}\right)}=0.1\left[-1,1,0.5\right]^{\top}$
and ${\rm b}_{2}^{\mathcal{B}\left({\rm R}\right)}=0.1\left[0,0,1\right]^{\top}$,
while the noise vectors $\omega_{1}^{\mathcal{B}\left({\rm R}\right)}$
and $\omega_{1}^{\mathcal{B}\left({\rm R}\right)}$ have zero mean
and ${\rm STD}=0.1\left({\rm m/sec}\right)$. The third inertial and
body-frame vectors are defined by ${\rm v}_{3}^{\mathcal{I}\left({\rm R}\right)}={\rm v}_{1}^{\mathcal{I}\left({\rm R}\right)}\times{\rm v}_{2}^{\mathcal{I}\left({\rm R}\right)}$
and ${\rm v}_{3}^{\mathcal{B}\left({\rm R}\right)}={\rm v}_{1}^{\mathcal{B}\left({\rm R}\right)}\times{\rm v}_{2}^{\mathcal{B}\left({\rm R}\right)}$.
It is worth noting that ${\rm v}_{i}^{\mathcal{B}\left({\rm R}\right)}$
and ${\rm v}_{i}^{\mathcal{I}\left({\rm R}\right)}$ are normalized
to $\upsilon_{i}^{\mathcal{B}\left({\rm R}\right)}$ and $\upsilon_{i}^{\mathcal{I}\left({\rm R}\right)}$,
respectively, for all $i=1,2,3$ using \eqref{eq:SE3STCH_Vector_norm}.
Hence, Assumption \ref{Assum:SE3STCH_1} holds. For the pose estimator
design presented in Subsection \ref{subsec:SE3_Passive-Filter}, $R_{y}$
is determined using SVD \cite{markley1988attitude}, for complete survey visit \cite{hashim2020AtiitudeSurvey}. The simulation
time is set to 25 seconds. Let us set the attitude estimate using
the angle-axis parameterization method outlined in \eqref{eq:SE3STCH_att_ang}
as $\hat{R}\left(0\right)=\mathcal{R}_{\alpha}\left(\alpha,u/\left\Vert u\right\Vert \right)$.
Define $\alpha=170\left({\rm deg}\right)$ and $u=\left[3,10,8\right]^{\top}$,
letting $||\tilde{R}\left(0\right)||_{{\rm I}}$ be very close to
the unstable equilibrium ($+1$) and setting the initial position
of the estimator as $\hat{P}\left(0\right)=\left[4,-3,5\right]^{\top}$.
The initial conditions are given below: {\small{}
	\[
	\boldsymbol{T}\left(0\right)=\mathbf{I}_{4},\hspace{1em}\hat{\boldsymbol{T}}\left(0\right)=\left[\begin{array}{cccc}
	-0.8816 & 0.2386 & 0.4074 & 4\\
	0.4498 & 0.1625 & 0.8782 & -3\\
	0.1433 & 0.9574 & -0.2505 & 5\\
	0 & 0 & 0 & 1
	\end{array}\right]
	\]
}{\small\par}

Design parameters and initial estimates are chosen as follows: $\gamma_{1}=1$,
$\gamma_{2}=1$, $k_{w}=8$, $k_{b}=0.1$, $k_{\sigma}=0.1$, $\varrho=0.2$,
$\hat{b}\left(0\right)=\mathbf{0}_{6\times1}$ and $\hat{\sigma}\left(0\right)=\mathbf{0}_{3\times1}$.
Also, the following color notation is adopted: black color describes
the true value, magenta refers to a measured value, red illustrates
the performance of the proposed nonlinear stochastic semi-direct pose
estimator (S-DIR), while blue demonstrates the performance of the
proposed nonlinear stochastic direct pose estimator (DIR).

\begin{figure}[h!]
	\centering{}\includegraphics[scale=0.26]{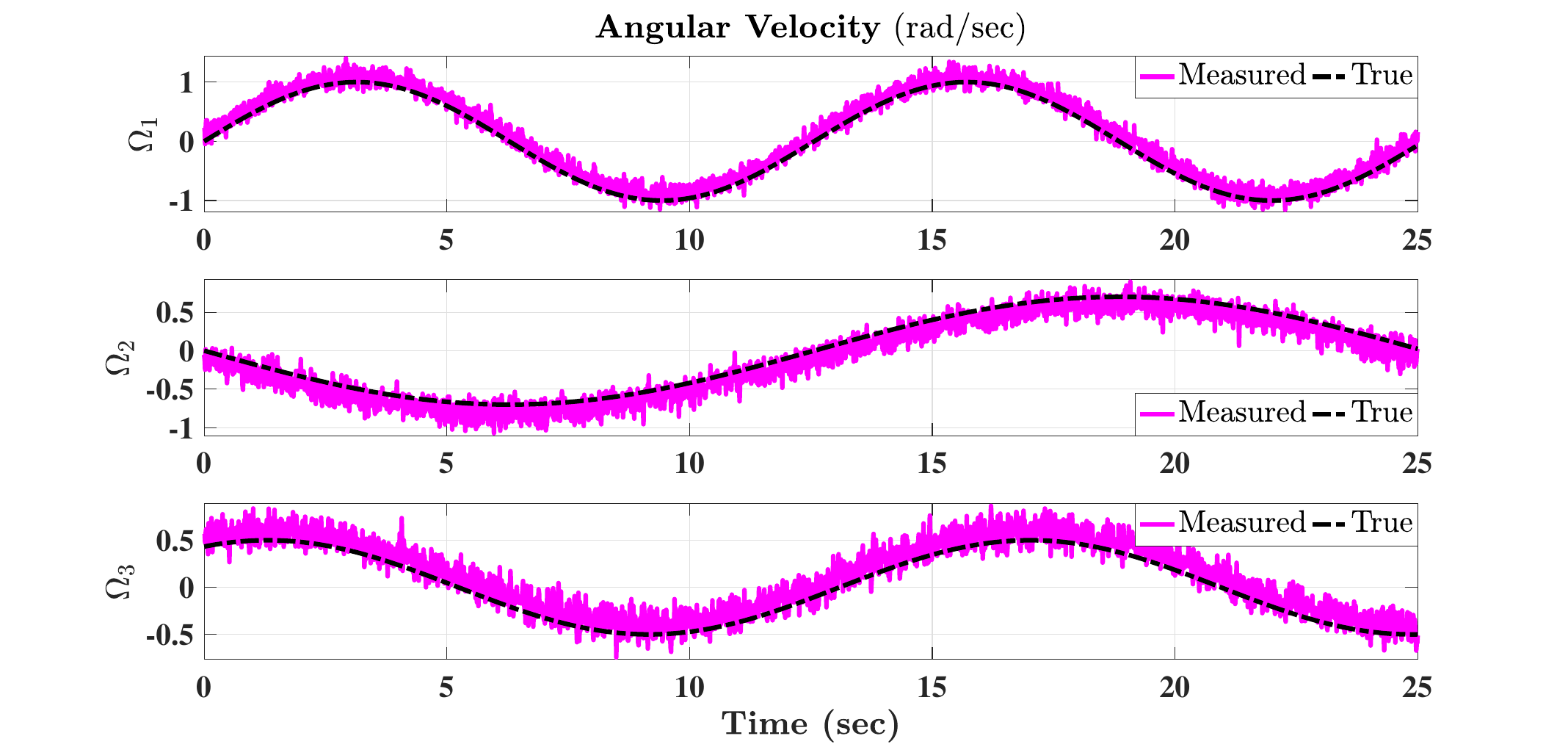}\caption{Angular Velocity: True vs Measured}
	\label{fig:SE3STOCH_Fig1_Ang} 
\end{figure}

\begin{figure}[h!]
	\centering{}\includegraphics[scale=0.26]{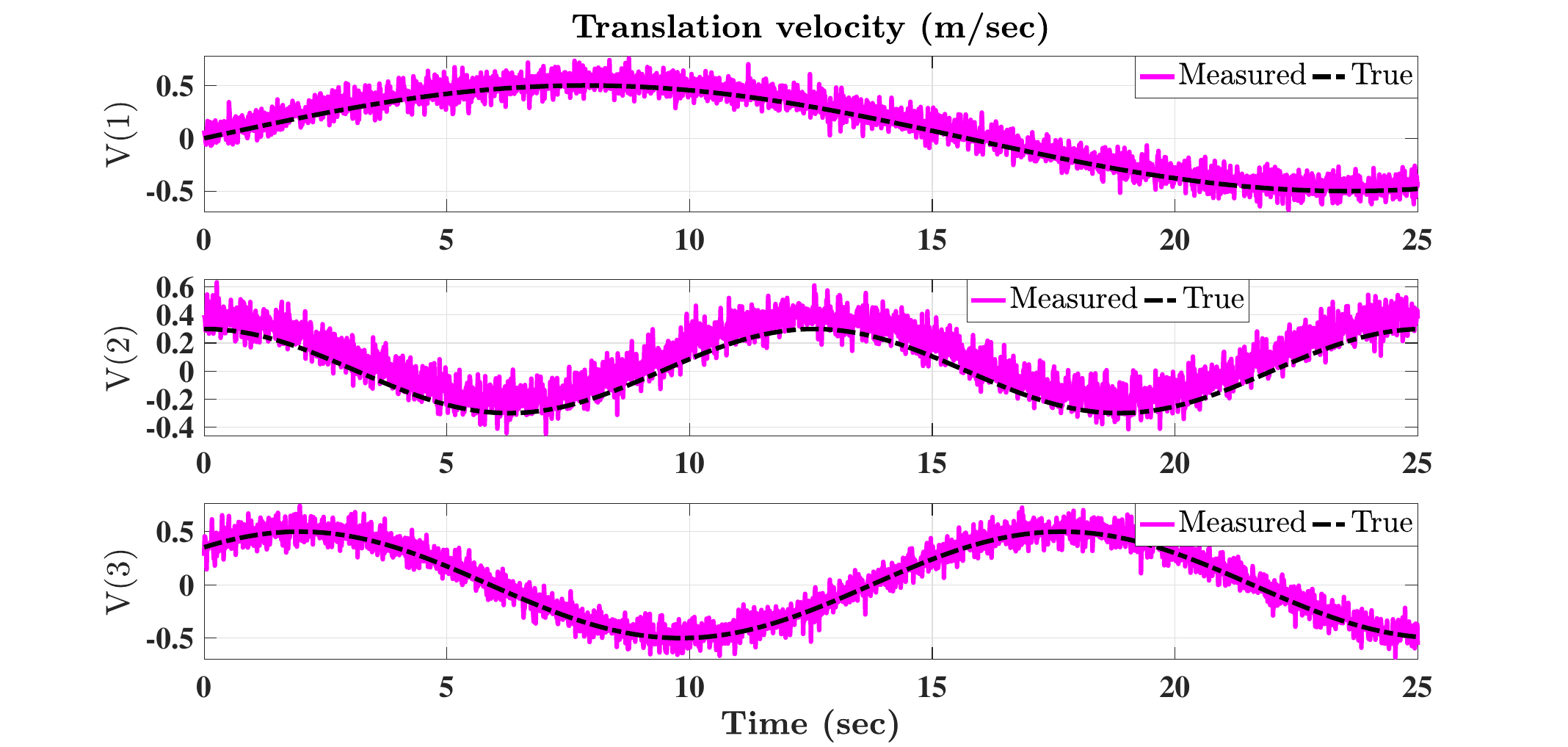}\caption{Translational Velocity: True vs Measured}
	\label{fig:SE3STOCH_Fig2_Vel} 
\end{figure}

\begin{figure}[h!]
	\centering{}\includegraphics[scale=0.26]{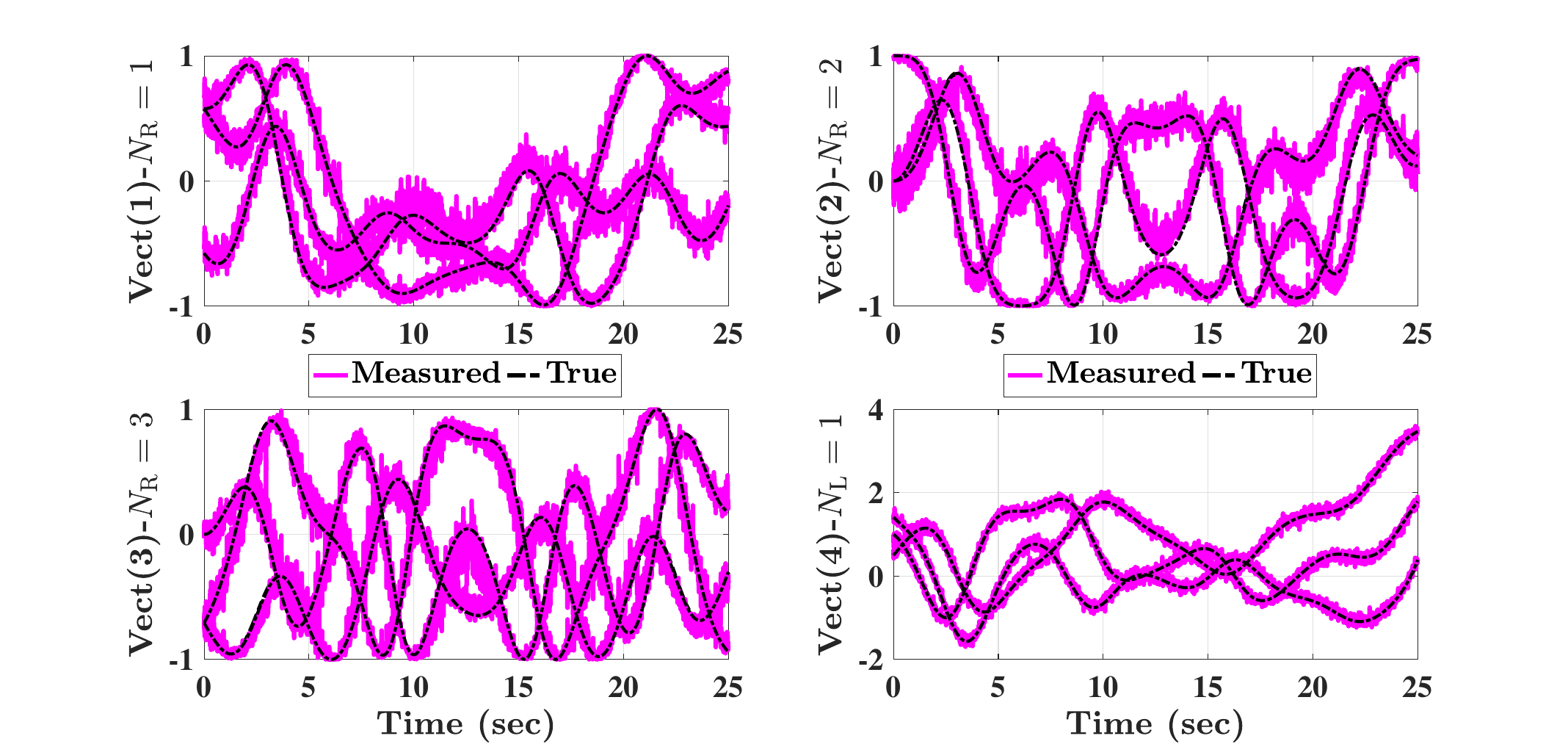}\caption{Body-frame data: True vs Measured}
	\label{fig:SE3STOCH_Fig3_Body} 
\end{figure}

\begin{figure}[h!]
	\centering{}\includegraphics[scale=0.26]{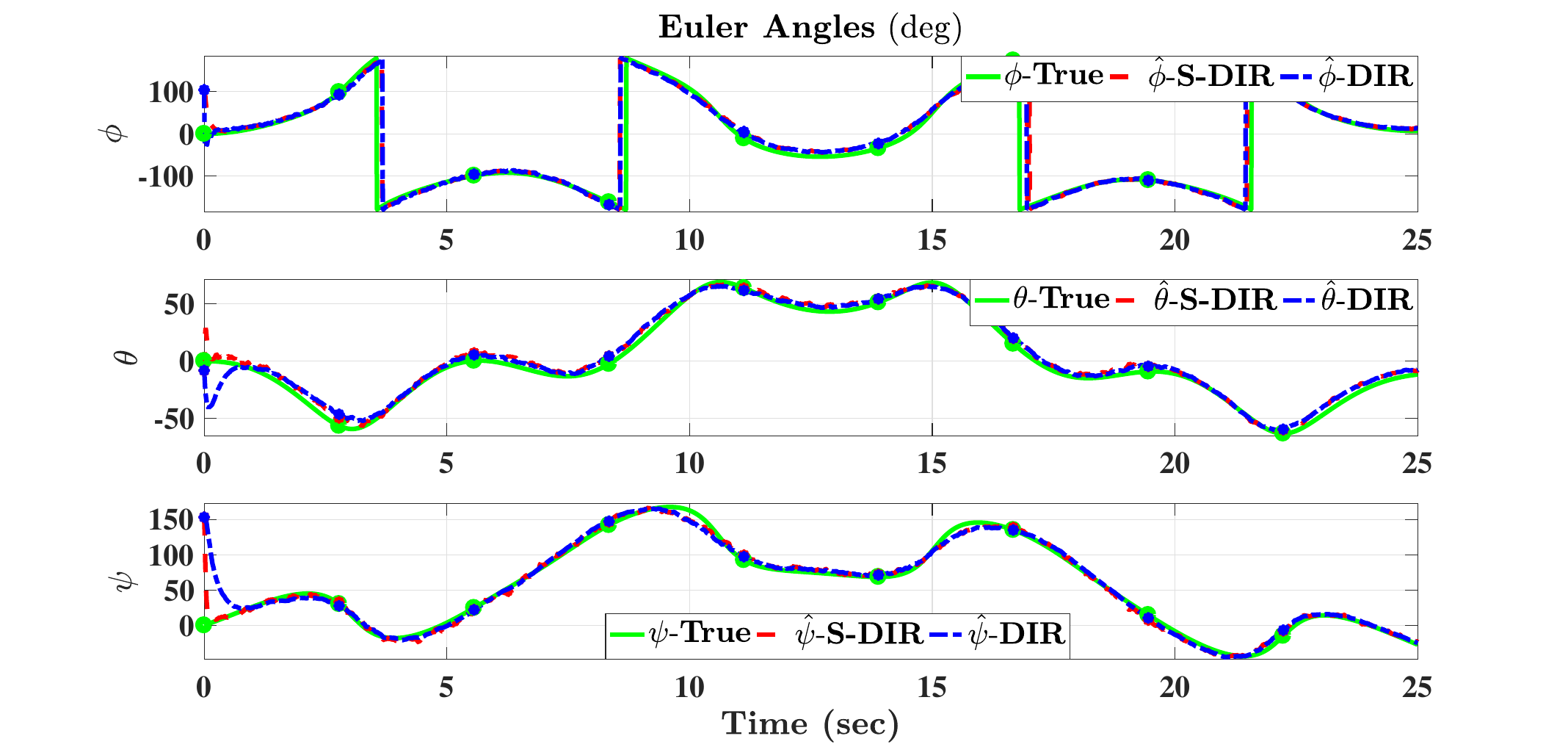}\caption{Euler angles: True vs proposed estimator.}
	\label{fig:SE3STOCH_Fig4_Euler} 
\end{figure}

\begin{figure}[h!]
	\centering{}\includegraphics[scale=0.26]{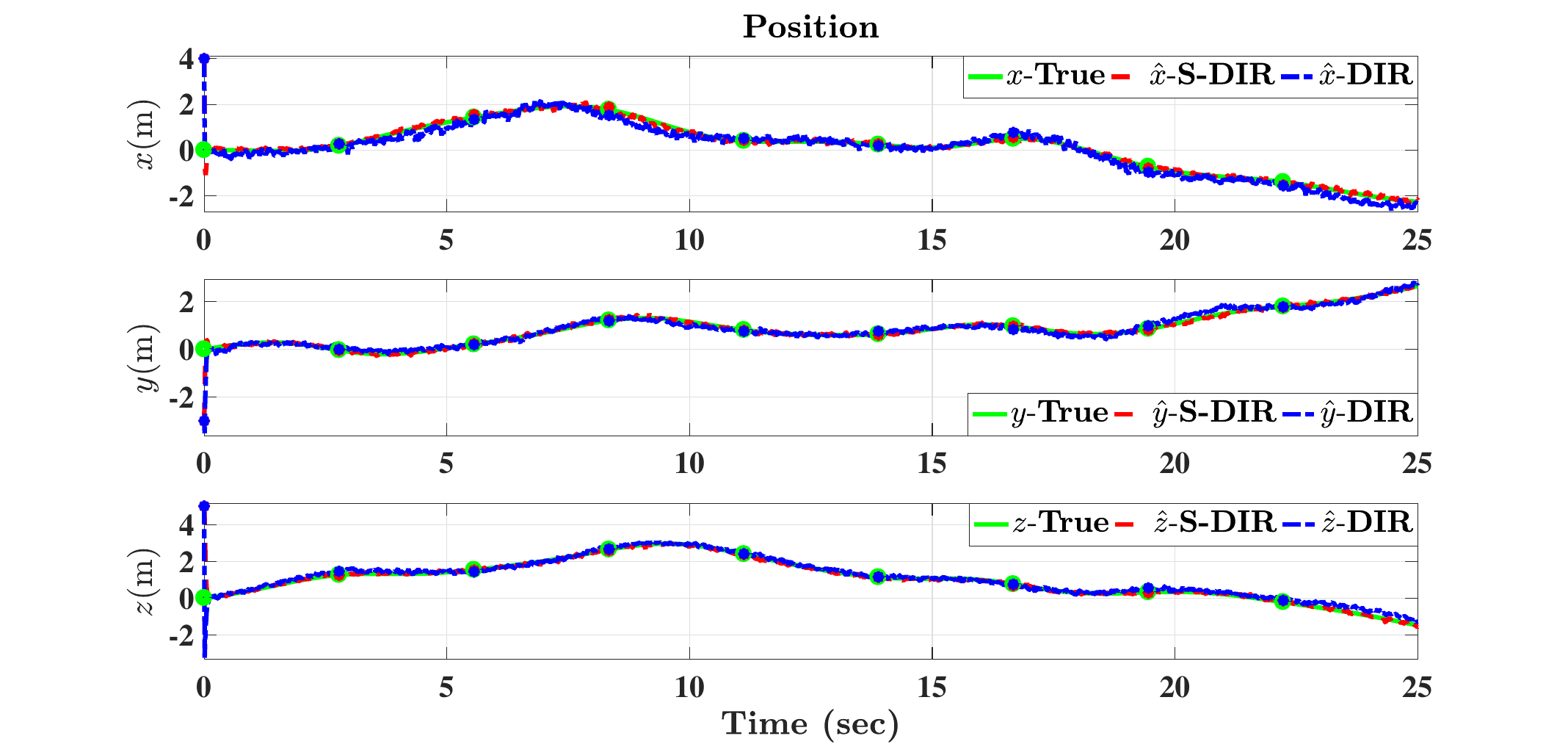}\caption{Rigid-body positions in 3D space: True vs proposed estimator.}
	\label{fig:SE3STOCH_Fig5_position} 
\end{figure}

\begin{figure}[h!]
	\centering{}\includegraphics[scale=0.26]{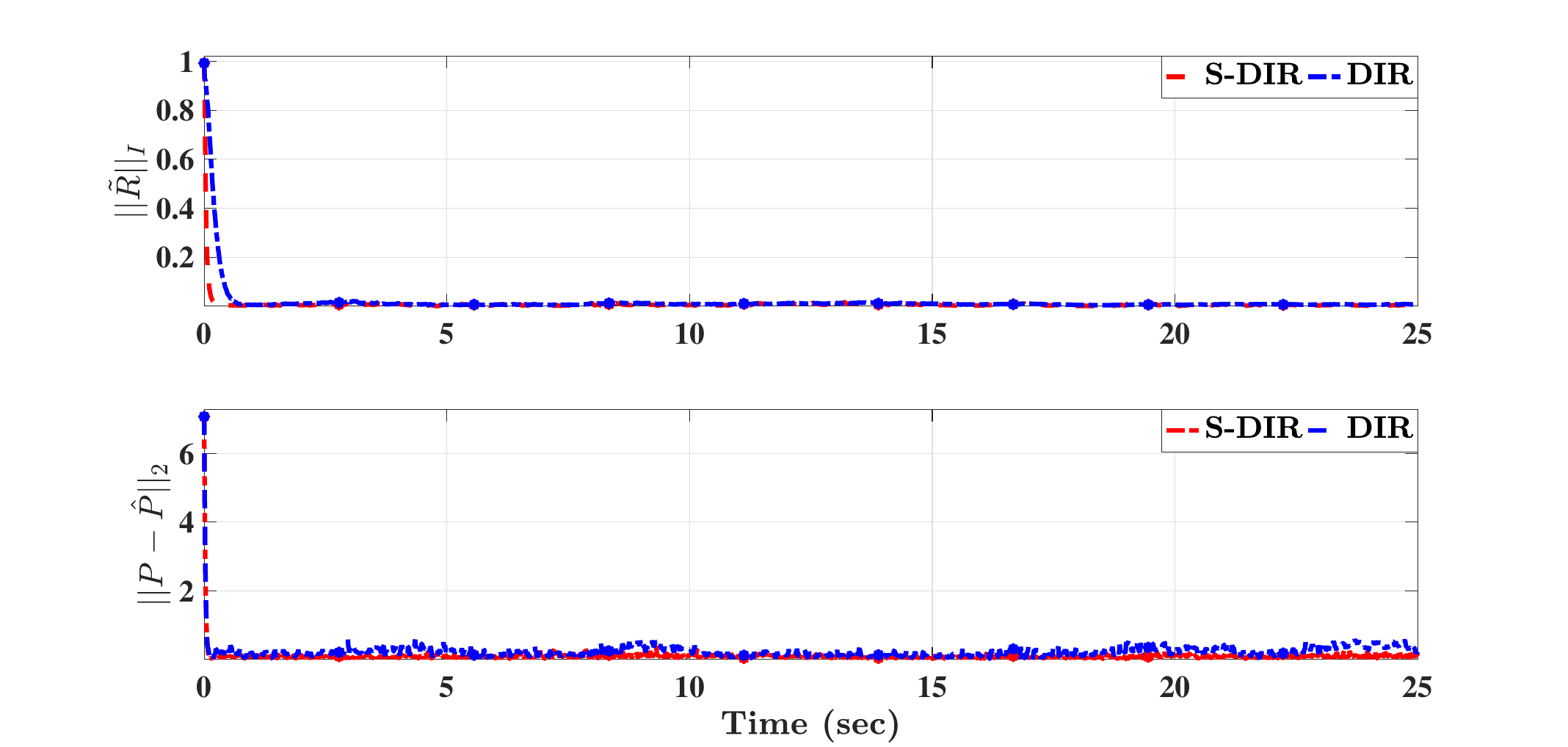}\caption{Tracking errors of $||\tilde{R}||_{{\rm I}}$ and $||P-\hat{P}||_{2}$. }
	\label{fig:SE3STOCH_Fig6_Error} 
\end{figure}

Fig. \ref{fig:SE3STOCH_Fig1_Ang}, \ref{fig:SE3STOCH_Fig2_Vel} and
\ref{fig:SE3STOCH_Fig3_Body} illustrate angular velocity, translational
velocity and body-frame vector measurements corrupted with high values
of bias and noise plotted against the true values. Fig. \ref{fig:SE3STOCH_Fig4_Euler}
demonstrates impressive tracking performance of the Euler angles which
are subject to large initialization error. Similarly, Fig. \ref{fig:SE3STOCH_Fig5_position}
depicts remarkable tracking performance of the rigid-body's position
in 3D space when large initialization error is present. Additionally,
the upper portion of Fig. \ref{fig:SE3STOCH_Fig6_Error} shows that%
{} $||\tilde{R}||_{{\rm I}}=\frac{1}{4}{\rm Tr}\{\mathbf{I}_{3}-\hat{R}R^{\top}\}$
initiated very close to the unstable equilibria approximated as (0.99)
and was regulated to the close proximity of the origin%
. In the same vein, the lower portion of Fig. \ref{fig:SE3STOCH_Fig6_Error}
demonstrates how $||P-\hat{P}||_{2}$ initiated at a high value and
steered to the close neighborhood of the origin%
. The impressive tracking performance presented in Fig. \ref{fig:SE3STOCH_Fig4_Euler},
\ref{fig:SE3STOCH_Fig5_position}, and \ref{fig:SE3STOCH_Fig6_Error}
illustrates the robustness of the proposed estimators against the
high values of bias, noise and initialization errors inherent to the
angular velocity, translational velocity, and body-frame vector measurements.

To compliment the estimator performance demonstrated in Fig. \ref{fig:SE3STOCH_Fig4_Euler},
\ref{fig:SE3STOCH_Fig5_position}, and \ref{fig:SE3STOCH_Fig6_Error}
with statistical analysis over the steady-state performance, Table
\ref{tab:SE3Filter_1} lists the mean and STD of $||\tilde{R}||_{{\rm I}}$
and $||P-\hat{P}||_{2}$ over the period of (8-25 sec) of the proposed
stochastic estimators%
. It can be noticed that both errors of the proposed stochastic estimators
exhibit small values of mean as well as STD which confirms the results
presented in Fig. \ref{fig:SE3STOCH_Fig4_Euler}, \ref{fig:SE3STOCH_Fig5_position},
and \ref{fig:SE3STOCH_Fig6_Error}. However, the semi-direct stochastic
pose estimator displays smaller mean and STD in comparison with the
direct stochastic pose estimator.
\begin{table}[h!]
	\caption{\label{tab:SE3Filter_1}Statistical analysis of $||\tilde{R}||_{{\rm I}}$
		and $||P^{e}||_{2}=||P-\hat{P}||_{2}$: proposed stochastic estimators.}
	
	\centering{}%
	\begin{tabular}{c|c|c||c|c|c||c}
		\hline 
		\multicolumn{7}{c}{Output data over the period (8-25 sec)}\tabularnewline
		\hline 
		\hline 
		Estimator & \multicolumn{3}{c|}{Stochastic (semi-direct)} & \multicolumn{3}{c}{Stochastic (direct)}\tabularnewline
		\hline 
		& $||\tilde{R}||_{I}$ & \multicolumn{2}{c|}{$||P^{e}||_{2}$} & $||\tilde{R}||_{I}$ & \multicolumn{2}{c}{$||P^{e}||_{2}$}\tabularnewline
		\hline 
		Mean & $0.005$ & \multicolumn{2}{c|}{$0.09$} & $0.008$ & \multicolumn{2}{c}{$0.227$}\tabularnewline
		\hline 
		STD & $0.0029$ & \multicolumn{2}{c|}{$0.046$} & $0.0024$ & \multicolumn{2}{c}{$0.125$}\tabularnewline
		\hline 
	\end{tabular}
\end{table}

Accordingly, the simulation results confirm the outstanding estimation
capability of the proposed stochastic pose estimators and their effectiveness
in handling uncertainty in the group velocity and body-frame vector
measurements as well as large initialization errors. The aforementioned
remarkable advantage makes the proposed stochastic estimators a perfect
match for uncertain data extracted from low-cost IMU and landmark
units. Although the semi-direct estimator has smaller values of $||\tilde{R}||_{{\rm I}}$
and $||P-\hat{P}||_{2}$ in comparison with the direct estimator,
it also requires pose reconstruction which in turn involves attitude
reconstruction using SVD \cite{markley1988attitude,hashim2018SO3Stochastic}.
Consequently, this adds complexity to the process and increases the
computational power requirements, in contrast to the direct stochastic
pose estimator which uses an available set of measurements directly. 

\section{Conclusion \label{sec:SE3_Conclusion}}

In this paper, the pose estimation problem has been addressed as a
nonlinear stochastic filtering problem on the Special Euclidean Group
$\mathbb{SE}\left(3\right)$. The group velocity vectors have been
assumed to be contaminated not only with unknown constant bias but
also with random Gaussian noise. Accordingly, two nonlinear stochastic
pose estimators on $\mathbb{SE}\left(3\right)$ have been proposed.
The closed loop error signals have been proven to be semi-globally
uniformly ultimately bounded in mean square. Simulation results and
statistical analysis revealed fast convergence capability of the proposed
estimators considering large initialized value of pose error and high
levels of unknown random noise and constant bias associated with velocity
measurements.

\section*{Acknowledgment}

The authors would like to thank \textbf{Maria Shaposhnikova} for proofreading
the article.

\section*{Appendix A \label{sec:SO3_PPF_STCH_AppendixA}}
\begin{center}
	\textbf{\large{}Proof of Lemma \ref{Lemm:SE3STCH_1}}{\large\par}
	\par\end{center}

Define the rotational matrix of a rigid-body in space by $R\in\mathbb{SO}\left(3\right)$.
Let $\rho\in\mathbb{R}^{3}$ be a Rodriguez parameters vector commonly
used for attitude representation \cite{shuster1993survey,hashim2019AtiitudeSurvey}. The mapping
from vector form to a 3-by-3 matrix $\mathcal{R}_{\rho}:\mathbb{R}^{3}\rightarrow\mathbb{SO}\left(3\right)$
is equivalent to
\begin{align}
\mathcal{R}_{\rho}\left(\rho\right)= & \frac{1}{1+||\rho||^{2}}\left(\left(1-||\rho||^{2}\right)\mathbf{I}_{3}+2\rho\rho^{\top}+2\left[\rho\right]_{\times}\right)\label{eq:SO3_PPF_STCH_SO3_Rodr}
\end{align}
Combining \eqref{eq:SO3_PPF_STCH_SO3_Rodr} and \eqref{eq:SE3STCH_Ecul_Dist}
one obtains
\begin{equation}
||R||_{{\rm I}}=||\rho||^{2}/(1+||\rho||^{2})\label{eq:SO3_PPF_STCH_TR2}
\end{equation}
For $\mathcal{R}_{\rho}=\mathcal{R}_{\rho}\left(\rho\right)$, the
anti-symmetric projection on the Lie-algebra of $\mathfrak{so}\left(3\right)$
is given by
\begin{align*}
\boldsymbol{\mathcal{P}}_{a}\left(R\right)=\frac{1}{2}\left(\mathcal{R}_{\rho}-\mathcal{R}_{\rho}^{\top}\right)= & 2\frac{1}{1+||\rho||^{2}}\left[\rho\right]_{\times}
\end{align*}
Accordingly, the vex of $\boldsymbol{\mathcal{P}}_{a}\left(R\right)$
is equivalent to
\begin{equation}
\mathbf{vex}\left(\boldsymbol{\mathcal{P}}_{a}\left(R\right)\right)=2\rho/(1+||\rho||^{2})\label{eq:SO3_PPF_STCH_VEX_Pa}
\end{equation}
Thus, using \eqref{eq:SO3_PPF_STCH_TR2} yields
\begin{equation}
\left(1-||R||_{{\rm I}}\right)||R||_{{\rm I}}=\frac{||\rho||^{2}}{\left(1+||\rho||^{2}\right)^{2}}\label{eq:SO3_PPF_STCH_append1}
\end{equation}
and \eqref{eq:SO3_PPF_STCH_VEX_Pa} shows that
\begin{equation}
||\mathbf{vex}\left(\boldsymbol{\mathcal{P}}_{a}\left(R\right)\right)||^{2}=4\frac{||\rho||^{2}}{\left(1+||\rho||^{2}\right)^{2}}\label{eq:SO3_PPF_STCH_append2}
\end{equation}
As such, \eqref{eq:SO3_PPF_STCH_append1} and \eqref{eq:SO3_PPF_STCH_append2}
justify \eqref{eq:SE3PPF_lemm1_1} in Lemma \ref{Lemm:SE3STCH_1}.
According to Subsection \ref{subsec:SE3_Direct-Filter} $\sum_{i=1}^{n}s_{i}=3$
in order to satisfy ${\rm Tr}\left\{ \mathbf{M}_{{\rm R}}\right\} =3$.
Consider $||R\mathbf{M}_{{\rm R}}||_{I}=\frac{1}{4}{\rm Tr}\left\{ \left(\mathbf{I}_{3}-R\right)\mathbf{M}_{{\rm R}}\right\} $
and the angle-axis parameterization in \eqref{eq:SE3STCH_att_ang}.
For $\mathbf{M}_{{\rm R}}=\left(\mathbf{M}_{{\rm R}}\right)^{\top}\in\mathbb{R}^{3}$,
one has ${\rm Tr}\left\{ \left[u\right]_{\times}\mathbf{M}_{{\rm R}}\right\} =0$
as given in \eqref{eq:SE3STCH_Identity5}. Thus, it could be found
that

\begin{align}
||R\mathbf{M}_{{\rm R}}||_{{\rm I}} & =\frac{1}{4}{\rm Tr}\left\{ -\left(\sin(\theta)\left[u\right]_{\times}+\left(1-\cos(\theta)\right)\left[u\right]_{\times}^{2}\right)\mathbf{M}_{{\rm R}}\right\} \nonumber \\
& =-\frac{1}{4}{\rm Tr}\left\{ \left(1-\cos(\theta)\right)\left[u\right]_{\times}^{2}\mathbf{M}_{{\rm R}}\right\} \label{eq:SO3_PPF_STCH_append3}
\end{align}
Accordingly, the following holds \cite{murray1994mathematical} 
\begin{equation}
||R||_{{\rm I}}=\frac{1}{2}\left(1-{\rm cos}\left(\theta\right)\right)={\rm sin}^{2}\left(\theta/2\right)\label{eq:SO3_PPF_STCH_append4}
\end{equation}
Hence, the unit axis vector is equivalent to \cite{shuster1993survey}
\[
u={\rm cot}\left(\theta/2\right)\rho
\]
Using $\left[u\right]_{\times}^{2}=-||u||^{2}\mathbf{I}_{3}+uu^{\top}$
in identity \eqref{eq:SE3STCH_Identity3}, one could rewrite the expression
in \eqref{eq:SO3_PPF_STCH_append3} as 
\begin{align*}
||R\mathbf{M}_{{\rm R}}||_{{\rm I}} & =\frac{1}{2}||R||_{{\rm I}}u^{\top}\bar{\mathbf{M}}_{{\rm R}}u\\
& =\frac{1}{2}||R||_{{\rm I}}{\rm cot}^{2}\left(\theta/2\right)\rho^{\top}\bar{\mathbf{M}}_{{\rm R}}\rho
\end{align*}
Based on \eqref{eq:SO3_PPF_STCH_append4}, ${\rm cos}^{2}\left(\theta/2\right)=1-||R||_{{\rm I}}$
such that 
\[
{\rm tan}^{2}\left(\theta/2\right)=\frac{||R||_{{\rm I}}}{1-||R||_{{\rm I}}}
\]
which means that $||R\mathbf{M}_{{\rm R}}||_{{\rm I}}$ formulated
in terms of $\rho$ is 
\begin{align}
||R\mathbf{M}_{{\rm R}}||_{{\rm I}} & =\frac{1}{2}\left(1-||R||_{I}\right)\rho^{\top}\bar{\mathbf{M}}_{{\rm R}}\rho=\frac{1}{2}\frac{\rho^{\top}\bar{\mathbf{M}}_{{\rm R}}\rho}{1+||\rho||^{2}}\label{eq:SO3_PPF_STCH_append_MBR_I}
\end{align}
Using \eqref{eq:SE3STCH_Identity1} and \eqref{eq:SE3STCH_Identity4},
the anti-symmetric projection operator of $R\mathbf{M}_{{\rm R}}$
is equivalent to
\begin{align*}
\boldsymbol{\mathcal{P}}_{a}\left(R\mathbf{M}_{{\rm R}}\right)= & \frac{\rho\rho^{\top}\mathbf{M}_{{\rm R}}-\mathbf{M}_{{\rm R}}\rho\rho^{\top}+\mathbf{M}_{{\rm R}}\left[\rho\right]_{\times}+\left[\rho\right]_{\times}\mathbf{M}_{{\rm R}}}{1+||\rho||^{2}}\\
= & \frac{\left[\left({\rm Tr}\left\{ \mathbf{M}_{{\rm R}}\right\} \mathbf{I}_{3}-\mathbf{M}_{{\rm R}}-\left[\rho\right]_{\times}\mathbf{M}_{{\rm R}}\right)\rho\right]_{\times}}{1+||\rho||^{2}}
\end{align*}
such that
\begin{align}
\mathcal{\mathbf{vex}}\left(\boldsymbol{\mathcal{P}}_{a}\left(R\mathbf{M}_{{\rm R}}\right)\right) & =\frac{(\mathbf{I}_{3}+\left[\rho\right]_{\times})}{1+||\rho||^{2}}\bar{\mathbf{M}}_{{\rm R}}\rho\label{eq:SO3_PPF_STCH_append_MBR_VEX}
\end{align}
One can verify that the 2-norm of the above result is
\begin{align*}
||\mathbf{vex}\left(\boldsymbol{\mathcal{P}}_{a}\left(R\mathbf{M}_{{\rm R}}\right)\right)||^{2} & =\frac{\rho^{\top}\bar{\mathbf{M}}_{{\rm R}}\left(\mathbf{I}_{3}-\left[\rho\right]_{\times}^{2}\right)\bar{\mathbf{M}}_{{\rm R}}\rho}{\left(1+||\rho||^{2}\right)^{2}}
\end{align*}
From identity \eqref{eq:SE3STCH_Identity3} $\left[\rho\right]_{\times}^{2}=-||\rho||^{2}\mathbf{I}_{3}+\rho\rho^{\top}$.
Thus, the following inequality holds
\begin{align}
||\mathbf{vex}\left(\boldsymbol{\mathcal{P}}_{a}\left(R\mathbf{M}_{{\rm R}}\right)\right)||^{2} & =\frac{\rho^{\top}\bar{\mathbf{M}}_{{\rm R}}\left(\mathbf{I}_{3}-\left[\rho\right]_{\times}^{2}\right)\bar{\mathbf{M}}_{{\rm R}}\rho}{\left(1+||\rho||^{2}\right)^{2}}\nonumber \\
& =\frac{\rho^{\top}\left(\bar{\mathbf{M}}_{{\rm R}}\right)^{2}\rho}{1+||\rho||^{2}}-\frac{\left(\rho^{\top}\bar{\mathbf{M}}_{{\rm R}}\rho\right)^{2}}{\left(1+||\rho||^{2}\right)^{2}}\nonumber \\
& \geq\underline{\lambda}\left(1-\frac{\left\Vert \rho\right\Vert ^{2}}{1+\left\Vert \rho\right\Vert ^{2}}\right)\frac{\rho^{\top}\bar{\mathbf{M}}_{{\rm R}}\rho}{1+||\rho||^{2}}\label{eq:SO3_PPF_STCH_append_VEX_ineq}
\end{align}
where $\left\Vert R\right\Vert _{{\rm I}}=\frac{\left\Vert \rho\right\Vert ^{2}}{1+\left\Vert \rho\right\Vert ^{2}}$,
and $\underline{\lambda}=\underline{\lambda}\left(\bar{\mathbf{M}}_{{\rm R}}\right)$
is the minimum singular value of $\bar{\mathbf{M}}_{{\rm R}}$. Since
$\mathbf{M}_{{\rm R}}$ has rank 3, one finds
\begin{align}
1-\left\Vert R\right\Vert _{{\rm I}} & ={\rm Tr}\{\frac{1}{12}\mathbf{I}_{3}+\frac{1}{4}R\}\nonumber \\
& ={\rm Tr}\{\frac{1}{12}\mathbf{I}_{3}+\frac{1}{4}R\mathbf{M}_{{\rm R}}\mathbf{M}_{{\rm R}}^{-1}\}\label{eq:SO3_PPF_STCH_append_rho2}
\end{align}
Based on \eqref{eq:SO3_PPF_STCH_append_VEX_ineq} and \eqref{eq:SO3_PPF_STCH_append_rho2},
the following inequality holds 
\begin{align*}
& ||\mathbf{vex}\left(\boldsymbol{\mathcal{P}}_{a}\left(R\mathbf{M}_{{\rm R}}\right)\right)||^{2}\\
& \hspace{5em}\geq\frac{\underline{\lambda}}{2}\left(1+{\rm Tr}\{R\mathbf{M}_{{\rm R}}\mathbf{M}_{{\rm R}}^{-1}\}\right)\left\Vert R\mathbf{M}_{{\rm R}}\right\Vert _{{\rm I}}
\end{align*}
which proves \eqref{eq:SE3PPF_lemm1_2} in Lemma \ref{Lemm:SE3STCH_1}.

\section*{Appendix B\label{sec:SO3_PPF_STCH_AppendixB} }
\begin{center}
	\textbf{\large{}{}{}{}{}{}{}{}{}{}{}{}Quaternion Representation}{\large{}{}{}
	} 
	\par\end{center}

\noindent Define $Q=[q_{0},q^{\top}]^{\top}\in\mathbb{S}^{3}$ as
a unit-quaternion with $q_{0}\in\mathbb{R}$ and $q\in\mathbb{R}^{3}$
such that $\mathbb{S}^{3}=\{\left.Q\in\mathbb{R}^{4}\right|||Q||=\sqrt{q_{0}^{2}+q^{\top}q}=1\}$.
$Q^{-1}=[\begin{array}{cc}
q_{0} & -q^{\top}\end{array}]^{\top}\in\mathbb{S}^{3}$ denotes the inverse of $Q$. Define $\odot$ as a quaternion product
where the quaternion multiplication of $Q_{1}=[\begin{array}{cc}
q_{01} & q_{1}^{\top}\end{array}]^{\top}\in\mathbb{S}^{3}$ and $Q_{2}=[\begin{array}{cc}
q_{02} & q_{2}^{\top}\end{array}]^{\top}\in\mathbb{S}^{3}$ is $Q_{1}\odot Q_{2}=[q_{01}q_{02}-q_{1}^{\top}q_{2},q_{01}q_{2}+q_{02}q_{1}+[q_{1}]_{\times}q_{2}]^{\top}$.
The mapping from unit-quaternion ($\mathbb{S}^{3}$) to $\mathbb{SO}\left(3\right)$
is described by $\mathcal{R}_{Q}:\mathbb{S}^{3}\rightarrow\mathbb{SO}\left(3\right)$
\begin{align}
\mathcal{R}_{Q} & =(q_{0}^{2}-||q||^{2})\mathbf{I}_{3}+2qq^{\top}+2q_{0}\left[q\right]_{\times}\in\mathbb{SO}\left(3\right)\label{eq:NAV_Append_SO3}
\end{align}
The quaternion identity is described by $Q_{{\rm I}}=[\pm1,0,0,0]^{\top}$
with $\mathcal{R}_{Q_{{\rm I}}}=\mathbf{I}_{3}$. For more information visit \cite{hashim2019AtiitudeSurvey}.  Define
the estimate of $Q=[q_{0},q^{\top}]^{\top}\in\mathbb{S}^{3}$ as $\hat{Q}=[\hat{q}_{0},\hat{q}^{\top}]^{\top}\in\mathbb{S}^{3}$
with $\mathcal{R}_{\hat{Q}}=(\hat{q}_{0}^{2}-||\hat{q}||^{2})\mathbf{I}_{3}+2\hat{q}\hat{q}^{\top}+2\hat{q}_{0}\left[\hat{q}\right]_{\times}$,
see the map in \eqref{eq:NAV_Append_SO3}. For any $x\in\mathbb{R}^{3}$
and $Q\in\mathbb{S}^{3}$, define the map
\begin{align*}
\overline{x} & =[0,x^{\top}]^{\top}\in\mathbb{R}^{4}\\
\overline{\mathbf{Y}(Q^{-1},x)} & =\left[\begin{array}{c}
0\\
\mathbf{Y}(Q^{-1},x)
\end{array}\right]=Q^{-1}\odot\left[\begin{array}{c}
0\\
x
\end{array}\right]\odot Q\\
\overline{\mathbf{Y}(Q,x)} & =\left[\begin{array}{c}
0\\
\mathbf{Y}(Q,x)
\end{array}\right]=Q\odot\left[\begin{array}{c}
0\\
x
\end{array}\right]\odot Q^{-1}
\end{align*}
The equivalent quaternion representation and complete implementation
steps of the filter in \eqref{eq:SE3PPF_Ty_dot}, \eqref{eq:SE3STCH_Wy_om},
\eqref{eq:SE3STCH_Wy_v}, \eqref{eq:SE3STCH_by_om}, \eqref{eq:SE3STCH_by_v},
and \eqref{eq:SE3STCH_sy_om} is:
\[
\begin{cases}
\upsilon_{i}^{\mathcal{B}} & =\mathbf{Y}(Q^{-1},\upsilon_{i}^{\mathcal{I}})\\
Q_{y} & :\text{Reconstructed by QUEST algorithm}\\
\tilde{Q} & =[\tilde{q}_{0},\tilde{q}^{\top}]^{\top}=\hat{Q}\odot Q_{y}^{-1}\\
\mathcal{E}_{R} & =1-\tilde{q}_{0}^{2}\\
P_{y} & =\frac{\sum_{i=1}^{N_{{\rm L}}}s_{i}^{{\rm L}}\left({\rm v}_{i}^{\mathcal{I}\left({\rm L}\right)}-\mathbf{Y}\left(Q_{y},{\rm v}_{i}^{\mathcal{B}\left({\rm L}\right)}\right)\right)}{\sum_{i=1}^{N_{{\rm L}}}k_{i}^{{\rm L}}}\\
\mathcal{E}_{P} & =\hat{P}-\mathbf{Y}\left(\tilde{Q},P_{y}\right)\\
\Gamma & =\Omega_{m}-\hat{b}-W\\
\dot{\hat{Q}} & =\frac{1}{2}\left[\begin{array}{cc}
0 & -\Gamma^{\top}\\
\Gamma & -\left[\Gamma\right]_{\times}
\end{array}\right]\hat{Q}\\
\dot{\hat{P}} & =\mathbf{Y}\left(\hat{Q},V_{m}-\hat{b}_{V}-W_{V}\right)\\
W_{\Omega} & =\frac{2\tilde{q}_{0}k_{w}}{1-\mathcal{E}_{R}}\left[\mathbf{Y}\left(\hat{Q}^{-1},\tilde{q}\right)\right]_{{\rm D}}\hat{\sigma}\\
W_{V} & =-\left[\mathbf{Y}\left(\hat{Q}^{-1},\hat{P}\right)\right]_{\times}W_{\Omega}+\frac{k_{w}}{\varrho}\mathbf{Y}\left(\hat{Q}^{-1},\mathcal{E}_{P}\right)\\
\dot{\hat{b}}_{\Omega} & =\gamma_{b}(1+\mathcal{E}_{R})\tilde{q}_{0}\exp(\mathcal{E}_{R})\mathbf{Y}\left(\hat{Q}^{-1},\tilde{q}\right)\\
& \hspace{1em}-\gamma_{b}\left\Vert \mathcal{E}_{P}\right\Vert ^{2}\left[\mathbf{Y}\left(\hat{Q}^{-1},\hat{P}\right)\right]_{\times}\mathbf{Y}\left(\hat{Q}^{-1},\mathcal{E}_{P}\right)\\
& \hspace{1em}-\gamma_{b}k_{b}\hat{b}_{\Omega}\\
\dot{\hat{b}}_{V} & =\gamma_{b}\left\Vert \mathcal{E}_{P}\right\Vert ^{2}\mathbf{Y}\left(\hat{Q}^{-1},\mathcal{E}_{P}\right)-\gamma_{b}k_{b}\hat{b}_{V}\\
K_{\mathcal{E}} & =\gamma_{\sigma}\frac{1+\mathcal{E}_{R}}{1-\mathcal{E}_{R}}\exp\left(\mathcal{E}_{R}\right)\\
\dot{\hat{\sigma}} & =4k_{w}\tilde{q}_{0}^{2}K_{\mathcal{E}}\left[\mathbf{Y}\left(\hat{Q}^{-1},\tilde{q}\right)\right]_{{\rm D}}\mathbf{Y}\left(\hat{Q}^{-1},\tilde{q}\right)\\
& \hspace{1em}-\gamma_{\sigma}k_{\sigma}\hat{\sigma}
\end{cases}
\]

\noindent The equivalent quaternion representation and complete implementation
steps of the filter in \eqref{eq:SE3PPF_Tvm_dot}, \eqref{eq:SE3STCH_Wvm_om},
\eqref{eq:SE3STCH_Wvm_v}, \eqref{eq:SE3STCH_bvm_om}, \eqref{eq:SE3STCH_bvm_v},
and \eqref{eq:SE3STCH_svm_om} is:
\[
\begin{cases}
\left[\begin{array}{c}
0\\
\upsilon_{i}^{\mathcal{B}}
\end{array}\right] & =\left[\begin{array}{c}
0\\
\mathbf{Y}(Q^{-1},\upsilon_{i}^{\mathcal{I}})
\end{array}\right]=Q^{-1}\odot\left[\begin{array}{c}
0\\
\upsilon_{i}^{\mathcal{I}}
\end{array}\right]\odot Q\\
\left[\begin{array}{c}
0\\
\hat{\upsilon}_{i}^{\mathcal{B}}
\end{array}\right] & =\left[\begin{array}{c}
0\\
\mathbf{Y}(\hat{Q}^{-1},\upsilon_{i}^{\mathcal{I}})
\end{array}\right]=\hat{Q}^{-1}\odot\left[\begin{array}{c}
0\\
\upsilon_{i}^{\mathcal{I}}
\end{array}\right]\odot\hat{Q}\\
\boldsymbol{\Upsilon} & =\mathcal{R}_{\hat{Q}}\sum_{i=1}^{N_{{\rm R}}}\left(\frac{s_{i}^{{\rm R}}}{2}\hat{\upsilon}_{i}^{\mathcal{B}\left({\rm R}\right)}\times\upsilon_{i}^{\mathcal{B}\left({\rm R}\right)}\right)\\
\mathcal{E}_{R} & =\frac{1}{4}\sum_{i=1}^{N_{{\rm R}}}\left(1-\left(\hat{\upsilon}_{i}^{\mathcal{B}\left({\rm R}\right)}\right)^{\top}\upsilon_{i}^{\mathcal{B}\left({\rm R}\right)}\right)\\
M_{1} & =\sum_{i=1}^{N_{{\rm R}}}s_{i}^{{\rm R}}\upsilon_{i}^{\mathcal{B}\left({\rm R}\right)}\left(\upsilon_{i}^{\mathcal{I}\left({\rm R}\right)}\right)^{\top}\\
M_{2} & =\left(\sum_{i=1}^{N_{{\rm R}}}s_{i}^{{\rm R}}\hat{\upsilon}_{i}^{\mathcal{B}\left({\rm R}\right)}\left(\upsilon_{i}^{\mathcal{I}\left({\rm R}\right)}\right)^{\top}\right)^{-1}\\
\mathcal{E}_{P} & =\tilde{P}=\hat{P}+\frac{1}{\mathbf{m}_{{\rm c}}}\left(\mathbf{Y}\left(\hat{Q},\mathbf{k}_{{\rm v}}\right)-M_{1}M_{2}\mathbf{m}_{{\rm v}}\right)\\
\Gamma & =\Omega_{m}-\hat{b}-W\\
\dot{\hat{Q}} & =\frac{1}{2}\left[\begin{array}{cc}
0 & -\Gamma^{\top}\\
\Gamma & -\left[\Gamma\right]_{\times}
\end{array}\right]\hat{Q}\\
\dot{\hat{P}} & =\mathbf{Y}\left(\hat{Q},V_{m}-\hat{b}_{V}-W_{V}\right)\\
W_{\Omega} & =\frac{4}{\underline{\lambda}_{1}}\frac{k_{w}}{1+{\rm Tr}\left\{ M_{1}M_{2}^{-1}\right\} }\left[\mathbf{Y}\left(\hat{Q}^{-1},\boldsymbol{\Upsilon}\right)\right]_{{\rm D}}\hat{\sigma}\\
W_{V} & =-\left[\mathbf{Y}\left(\hat{Q}^{-1},\hat{P}\right)\right]_{\times}W_{\Omega}+\frac{k_{w}}{\varrho}\mathbf{Y}\left(\hat{Q}^{-1},\mathcal{E}_{P}\right)\\
\dot{\hat{b}}_{\Omega} & =\frac{\gamma_{b}}{2}\left(1+\mathcal{E}_{R}\right)\exp\left(\mathcal{E}_{R}\right)\mathbf{Y}\left(\hat{Q}^{-1},\boldsymbol{\Upsilon}\right)\\
& \hspace{1em}-\gamma_{b}\left\Vert \mathcal{E}_{P}\right\Vert ^{2}\left[\mathbf{Y}\left(\hat{Q}^{-1},\hat{P}\right)\right]_{\times}\mathbf{Y}\left(\hat{Q}^{-1},\mathcal{E}_{P}\right)\\
&\hspace{1em}-\gamma_{b}k_{b}\hat{b}_{\Omega}\\
\dot{\hat{b}}_{V} & =\gamma_{b}\left\Vert \mathcal{E}_{P}\right\Vert ^{2}\mathbf{Y}\left(\hat{Q}^{-1},\mathcal{E}_{P}\right)-\gamma_{b}k_{b}\hat{b}_{V}\\
K_{\mathcal{E}} & =\gamma_{\sigma}\frac{1+\mathcal{E}_{R}}{1+{\rm Tr}\left\{ M_{1}M_{2}^{-1}\right\} }\exp\left(\mathcal{E}_{R}\right)\\
\dot{\hat{\sigma}} & =\frac{2k_{w}}{\underline{\lambda}_{1}}K_{\mathcal{E}}\left[\mathbf{Y}\left(\hat{Q}^{-1},\boldsymbol{\Upsilon}\right)\right]_{{\rm D}}\mathbf{Y}\left(\hat{Q}^{-1},\boldsymbol{\Upsilon}\right)\\
&\hspace{1em}-\gamma_{\sigma}k_{\sigma}\hat{\sigma}
\end{cases}
\]

\bibliographystyle{IEEEtran}
\bibliography{bib_SE3_STOCH}


\end{document}